\documentclass[sigconf]{acmart}

\usepackage{balance}
\usepackage{graphics}
\usepackage{hyperref}
\usepackage{color}
\usepackage{booktabs}
\usepackage{textcomp}
\usepackage{subcaption}
\usepackage{enumerate}
\usepackage{xcolor}
\usepackage{lipsum}
\usepackage{makecell}
\usepackage{multicol}
\usepackage{multirow}
\usepackage{array}
\usepackage{verbatimbox}
\usepackage{enumitem}
\usepackage{amsmath}
\usepackage{stfloats}
\usepackage{graphicx}
\usepackage{amsthm}
\usepackage{listings}
\usepackage{caption} 
\usepackage{xspace}
\usepackage[linesnumbered]{algorithm2e}
\usepackage{algpseudocode}
\usepackage{tabularx}
\usepackage{arydshln}
\usepackage[bottom]{footmisc}
\usepackage{stackengine}
\usepackage{placeins}
\usepackage{bbding}

\newcommand*{\eg}{\textit{e.g.},\xspace}
\newcommand*{\ie}{\textit{i.e.},\xspace}
\newcommand*{\vs}{\textit{vs.}\xspace}

\newcommand*{\etal}{\textit{et~al.}\xspace}

\newcolumntype{L}[1]{>{\raggedright\let\newline\\\arraybackslash\hspace{0pt}}m{#1}}
\newcolumntype{C}[1]{>{\centering\let\newline\\\arraybackslash\hspace{0pt}}m{#1}}
\newcolumntype{R}[1]{>{\raggedleft\let\newline\\\arraybackslash\hspace{0pt}}m{#1}}

\newcommand\pquote[2]{{``\textit{#2}'' (P#1)}}

\algnewcommand{\IfThenElse}[3]{
  \State \algorithmicif\ #1\ \algorithmicthen\ #2\ \algorithmicelse\ #3}

\definecolor{downredcolor}{HTML}{e31a1c}
\definecolor{upgreencolor}{HTML}{33a02c}

\definecolor{DarkGreen}{HTML}{5DAC81}

\newcommand\projectname{Time2Stop}

\AtBeginDocument{%
  \providecommand\BibTeX{{%
    \normalfont B\kern-0.5em{\scshape i\kern-0.25em b}\kern-0.8em\TeX}}}

\copyrightyear{2024}
\acmYear{2024}
\setcopyright{acmlicensed}\acmConference[CHI '24]{Proceedings of the CHI Conference on Human Factors in Computing Systems}{May 11--16, 2024}{Honolulu, HI, USA}
\acmBooktitle{Proceedings of the CHI Conference on Human Factors in Computing Systems (CHI '24), May 11--16, 2024, Honolulu, HI, USA}
\acmDOI{10.1145/3613904.3642747}
\acmISBN{979-8-4007-0330-0/24/05}

\usepackage{acmart-taps}
\begin{document}

\title{\projectname{}: Adaptive and Explainable Human-AI Loop for Smartphone Overuse Intervention}

\author{Adiba Orzikulova}
\affiliation{%
  \institution{KAIST}
  \city{} \state{} \country{Republic of Korea}
}
\email{adiorz@kaist.ac.kr}

\author{Han Xiao}
\affiliation{%
  \institution{Beijing University of Posts and Telecommunications}
  \city{} \state{} \country{Beijing, Beijing, China}
}
\email{umihara@bupt.edu.cn}

\author{Zhipeng Li}
\affiliation{%
  \institution{Tsinghua University}
  \city{Beijing} \state{} \country{China}
}
\email{lizhipeng0603@gmail.com}

\author{Yukang Yan}
\affiliation{%
  \institution{Carnegie Mellon University}
  \city{} \state{} \country{Pittsburgh, Pennsylvania, USA}
}
\email{yukangy@andrew.cmu.edu}

\author{Yuntao Wang}
\authornote{Corresponding Author for Funding}
\affiliation{%
  \institution{Tsinghua University}
  \city{} \state{} \country{Beijing, Beijing, China}
}
\email{yuntaowang@tsinghua.edu.cn}

\author{Yuanchun Shi}
\affiliation{%
  \institution{Tsinghua University}
  \city{} \state{} \country{Beijing, Beijing, China}
}
\email{shiyc@tsinghua.edu.cn}

\author{Marzyeh Ghassemi}
\affiliation{%
  \institution{Massachusetts Institute of Technology}
  \city{} \state{} \country{Cambridge, MA, USA}
}
\email{mghassem@mit.edu}

\author{Sung-Ju Lee}
\affiliation{%
  \institution{KAIST}
  \city{} \state{} \country{Republic of Korea}
}
\email{profsj@kaist.ac.kr}

\author{Anind K. Dey}
\affiliation{%
  \institution{University of Washington}
  \city{} \state{} \country{Seattle, UW, USA}
}
\email{anind@uw.edu}

\author{Xuhai Xu}
\authornote{Corresponding Author}
\affiliation{%
  \institution{Massachusetts Institute of Technology}
  \city{} \state{} \country{Cambridge, MA, USA}
}
\email{xoxu@mit.edu}

\renewcommand{\shortauthors}{Adiba Orzikulova et al.}
\renewcommand{\shorttitle}{\projectname{}}

\begin{abstract}
Despite a rich history of investigating smartphone overuse intervention techniques, AI-based just-in-time adaptive intervention (JITAI) methods for overuse reduction are lacking. We develop \projectname{}, an intelligent, adaptive, and explainable JITAI system that leverages machine learning to identify optimal intervention timings, introduces interventions with transparent AI explanations, and collects user feedback to establish a human-AI loop and adapt the intervention model over time. We conducted an 8-week field experiment (N=71) to evaluate the effectiveness of both the adaptation and explanation aspects of \projectname{}.
Our results indicate that our adaptive models significantly outperform the baseline methods on intervention accuracy (>32.8\%  relatively) and receptivity (>8.0\%). 
In addition, incorporating explanations further enhances the effectiveness by 53.8\% and 11.4\% on accuracy and receptivity, respectively. Moreover, \projectname{} significantly reduces overuse, decreasing app visit frequency by 7.0$\sim$8.9\%. Our subjective data also echoed these quantitative measures. Participants preferred the adaptive interventions and rated the system highly on intervention time accuracy, effectiveness, and level of trust.
We envision our work can inspire future research on JITAI systems with a human-AI loop to evolve with users.

\end{abstract}



\maketitle

\section{Introduction}
\label{sec:introduction}

\begin{figure*}
    \centering
    \includegraphics[width=\textwidth]{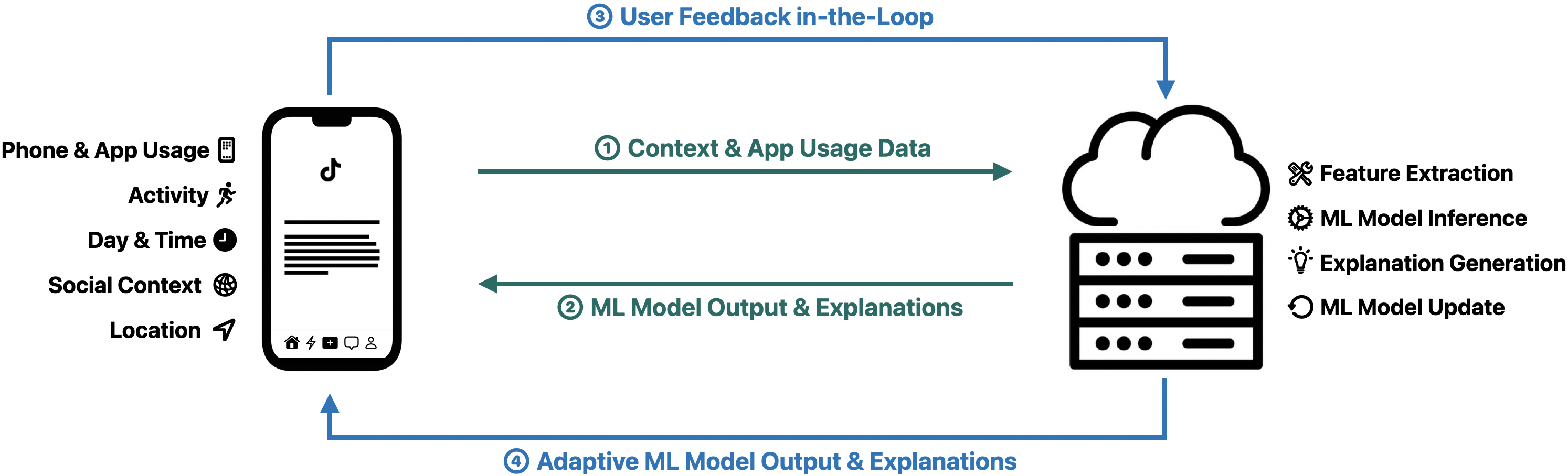}
    \caption{\projectname{} System Overview. The overall interaction flow consists of two loops. The first loop (green) includes: \textcircled{1} The mobile app continuously gathers contextual and app usage data (left) and transmits them to the cloud server. \textcircled{2} On the cloud server's end, feature extraction, ML model inference, and explanation generation occur (right). The ML model output and explanations are sent back to the user.
    The second loop (loop) includes: \textcircled{3} In cases where the model predicts ``overuse'', an intervention would show up while allowing users to provide feedback. The feedback is then forwarded to the cloud server to update the ML model. \textcircled{4} The updated ML model is subsequently employed to provide more personalized and adaptive interventions.}
    \label{fig:system-overview}
\end{figure*}

The rapid advancement of technology has empowered the use of mobile devices to engage in almost every aspect of our lives. While bringing us convenience, smartphones also introduce numerous potential risks~\cite{fischer2019risk,parasuraman2017smartphone}.
Smartphone overuse is considered a major social problem as it adversely affects individuals' physical health~(\eg headaches~\cite{demirci2016headache}, chronic neck pain~\cite{zhuang2021association}, sleep disturbance~\cite{lapointe2013is}); mental well-being (\eg anxiety and depression~\cite{hartanto2016smartphone, bevan2012negative}, impaired cognitive abilities~\cite{yasin2022digital}); and social wellness~(\eg distraction~\cite{lyngs2019self}, family conflicts~\cite{turel2008blackberry}, degradation of academic and work performance~\cite{giunchiglia2018mobile}).

A plethora of research has been invested in designing and experimenting with various digital intervention tools and techniques to regulate smartphone overuse. These mechanisms promote digital well-being by informing users about their usage statistics~\cite{digitalWellbeing, screenTime, monge2019race, lyngs2022goldilocks}, restricting access to distracting apps~\cite{xu_typeout_2022, kim2019lockntype, ko2016lock, lochtefeld2013appdetox} or app functionalities~\cite{cho_reflect_2021, orzikulova2023finerme, lukoff2021design}.
While the proposed mechanisms were beneficial in enhancing self-awareness and reducing smartphone usage time, they primarily intervened based on simple criteria, \eg upon opening a specific app, at pre-determined intervals, or after achieving daily usage goals.
However, due to the considerable variability of human behavior, interventions based on these basic criteria may not be optimal.
For instance, users sometimes need to take a break, and blocking usage without considering such contexts could lead to sub-optimal designs and impact user experience. A system should offer intelligent interventions tailored to the user's preferences, app characteristics, dynamically changing context, and individual usage patterns.

In mobile health, Just-In-Time Adaptive Intervention (JITAI) was introduced as a promising technique that provides appropriate support at opportune times while dynamically adapting to users' internal and external context~\cite{nahum-shani_just--time_2018, nahum-shani_translating_2021}.
Traditional JITAI-driven intervention systems incorporate a predefined set of rules (such as users' location) to determine the delivery time or content~\cite{riley2008internet,gustafson2014smartphone,lukoff2023switchTube}.
There has been initial research leveraging artificial intelligence (AI) and machine learning (ML) to deliver interventions, as AI can analyze large amounts of data and identify patterns that might not be captured through manual rule-setting~\cite{mishra_detecting_2021,kunzler2019exploring}.
Furthermore, in a human-AI-loop setup, users can offer feedback to the AI, allowing the model to enhance its predictions and personalize interventions based on individual needs and behaviors. 

However, very little work explores empowering AI-based JITAI with a human-in-the-loop setup~\cite{kim_prediction_2022,liao_personalized_2020}.
There is no prior work leveraging AI-based JITAI in the realm of smartphone overuse, not to mention the human-in-the-loop setup.
Employing JITAI-based interventions for smartphone overuse is challenging as it requires a real-time ML pipeline (reacting in a few seconds when the user enters an app) and prompt adaptability to users' constantly evolving habits (updating the model on a daily basis).

Despite the accuracy of black-box AI models, they often face challenges of interpretability and transparency. This gave rise to the recent advance of Explainable AI (XAI) to help users comprehend AI systems' decisions, thereby fostering user trust and collaboration with AI~\cite{ribeiro_why_2016,barredo_arrieta_explainable_2020}. Recent intervention systems employed XAI to personalize education~\cite{hur2022using}, manage stress~\cite{kim_prediction_2022}, and set fitness goals~\cite{wozniak2020exploring}.
However, there is no prior work integrating XAI into JITAI-based smartphone intervention. It can improve intervention delivery transparency, handle confusion caused by unexpected interventions, and cultivate users' trust through human-AI interaction.

No prior work has used AI-driven JITAI for smartphone overuse nor incorporated a human-in-the-loop setup. Moreover, integrating XAI into JITAI-based smartphone interventions remains unexplored. 
To address these gaps, we design and implement \projectname{}, \textbf{an intelligent, adaptive, and explainable smartphone overuse intervention system grounded in JITAI principles while taking user feedback in the loop}.
Our system consists of four major parts (see Figure~\ref{fig:system-overview}): (1) a smartphone-based sensing app to collect users' contexts and behavior, (2) a cloud-based ML pipeline that extracts behavior features, detects potential smartphone overuse behavior, and generates explanations, (3) an interface on local devices that introduces interventions when the ML model detects overuse behavior, provides intervention explanations, and collects user-provided feedback (\eg users' opinions on the accuracy of the intervention), and (4) a human-AI feedback loop that leverage users' reactions to update the ML model.

We conducted an eight-week field experiment (N=71) to deploy and evaluate the effectiveness of the two major characteristics of \projectname{}:
(a)~\textit{Adaptive}, updating the model based on user feedback in the human-AI loop; 
(b)~\textit{Explainable}, providing feature explanations based on user behavior and model outcomes.
Our findings demonstrate that interventions with the \textit{adaptive} models significantly outperform both the basic (statistics-based) and the personalized ML (but non-adaptive) methods on smartphone overuse prediction accuracy (32.8$\sim$55.5\% relative advantage) and intervention receptivity (8.0$\sim$29.0\%).
Moreover, incorporating \textit{explanations} in interventions can further enhance the effectiveness (53.8$\sim$97.5\% on relative accuracy, 11.4$\sim$39.6\% on relative receptivity).
From the perspective of smartphone usage behavior, our results indicate that app visit frequency was reduced significantly with the help of \projectname{} (7.0$\sim$8.9\%).
We also observe an interesting nuance of explanations' effect on user behavior.
Our qualitative results from the exit questionnaire and interview align with the quantitative findings, further supporting the advantage of \projectname{}. 
We discuss the mixed effects of explanations and the design considerations and ethical concerns of AI-based interventions.

The main contributions of our paper can be summarized as follows:
\begin{itemize}
    
    \item We designed and implemented \projectname{}, an adaptive and explainable JITAI-grounded intervention system for smartphone overuse. \projectname{} performs real-time inference on smartphone overuse behavior, introduces just-in-time intelligent intervention with explanations, and evolves based on users' feedback.
   
    \item We conducted a longitudinal field experiment with micro-randomized trials to demonstrate the effectiveness of empowering interventions to be adaptive and explainable. Our results show that \projectname{} significantly outperforms baseline techniques.
    \item We share the lessons learned, and discuss the design considerations and ethical concerns when creating AI-based smartphone intervention systems with humans in the loop.
\end{itemize}

We envision that empowering AI-based JITAI with both human-in-the-loop and AI explanations can go beyond smartphone overuse. When focusing on another application, careful design of the human-AI-loop (\eg updating models with user feedback in our case) and the integration of an appropriate level of explanation (\eg highlighting feature types in our case) is necessary.

\section{Background}
\label{sec:background}
We first summarize existing research in smartphone overuse intervention techniques and just-in-time adaptive intervention~(JITAI) methods. We also provide a brief overview of explainable AI~(XAI), and emerging research in the intersection of XAI and JITAI intervention domains.

\subsection{Smartphone Overuse and Intervention}
\label{sub:background:overuse}
Excessive smartphone usage has been connected to a variety of undesirable effects, such as distraction~\cite{lyngs2019self}, anxiety and depression~\cite{hartanto2016smartphone}, neck pain~\cite{zhuang2021association}, and disruptions in sleep patterns~\cite{lapointe2013is}. 
In response to the negative effects of smartphone overuse, there has been a wide range of commercial products and research solutions. For example, ScreenTime~\cite{screenTime} on iOS and Digital Wellbeing~\cite{digitalWellbeing} on Android are built-in tools designed to assist users in tracking app usage and setting usage limits. In addition, there are also various third-party apps for overuse intervention, such as 
Forest~\cite{forestApp}, Digital Detox~\cite{digitalDetoxApp}, and StayFree~\cite{stayFreeApp}.

Within the academic sphere, researchers have proposed a large array of works in the smartphone overuse intervention domain~\cite{kovacs2019habitlab, monge2019race, cho_reflect_2021, orzikulova2023finerme, lukoff2021design,lu2024interactout}.
These methods can be generally divided into two categories: (1) sending notifications or reminders~\cite{hiniker2016mytime,NUGU}, and (2) blocking user access to apps or phones~\cite{xu_typeout_2022, kim2019lockntype, ko2016lock}.
The first category aims to softly persuade users to limit digital consumption. For example, 
MyTime~\cite{hiniker2016mytime} informs users about their usage time and sends notifications upon reaching their time limit.
NUGU \cite{NUGU} leveraged social effects by visualizing smartphone usage among social groups via a scoreboard.
The second category is more restrictive, aiming to introduce a higher interaction cost and a gulf of instant gratification. For instance, 
LockNType~\cite{kim2019lockntype} adds a typing task before users can access their apps to trigger System 2, \ie the reasoning and analytical system in the Dual Process Theory~\cite{hofmann_impulse_2009}.
Building on top of this work, TypeOut~\cite{xu_typeout_2022} integrates the typing task with self-affirmation to effectively mitigate smartphone overuse.

Other than dividing interventions based on their restrictiveness, another line of work was devoted to building smartphone interventions at different granularities: device-level, app-level, and feature-level.
A study by Roffarello \etal~\cite{monge2019race} found that intervening at the app-level is more effective compared to device-level, as the former can generate more precise and interpretable statistics for users. Orzikulova ~\etal~\cite{orzikulova2023finerme}  investigated app-level (\eg restricting Instagram and YouTube apps) and feature-level (\eg limiting the usage of app features such as viewing suggested feed on Instagram and watching shorts on YouTube) interventions on mobile social media apps. The results indicated that feature-level restrictive interventions were particularly effective in reducing the time spent on passive phone usage (\eg watching short videos).

While these intervention techniques are beneficial in enhancing user awareness and reducing phone use time, they rely on basic conditions (such as being triggered upon opening an app) or simple parameters specified by users (such as the daily usage limit). 
However, users' behavior is changing dynamically and these manual rules are often outdated.
For an intelligent smartphone intervention system, it is essential to account for user's preferences, contexts, and smartphone usage patterns, so that it can achieve a good intervention performance continuously.
To address this gap, \projectname{} implements real-time adaptability to accommodate users' evolving contexts and behavior.

\subsection{Just-in-Time Adaptive Interventions}\label{sub:background:jitai}
In the context of mobile health, JITAI is an emerging intervention design methodology that seeks to deliver tailored and timely support by dynamically adjusting to an individual's internal and contextual conditions~\cite{nahum-shani_just--time_2018, nahum-shani_translating_2021}. For a JITAI to be effective, intervention needs to be delivered when the user is both \textit{vulnerable} and \textit{receptive}~\cite{nahum-shani_just--time_2018, mishra2021detecting}. Vulnerability denotes a period during which individuals are more susceptible to experiencing negative health consequences (\ie overusing smartphones in our case), whereas receptivity pertains to their ability to receive and process provided interventions (\ie accepting intervention and stopping using phones).

JITAI-driven systems may either be rule-based~\cite{gustafson2014smartphone, riley2008internet} or AI-based~\cite{mishra_detecting_2021, kunzler2019exploring, kim_prediction_2022}.
Rule-based JITAI relies on predefined sets of rules and conditions to trigger interventions, typically established by domain experts.
For example, Gustafson~\etal~\cite{gustafson2014smartphone} designed a JITAI system for alcohol consumption that will trigger intervention when users approach high-risk locations such as bars.
Lukoff~\etal~\cite{lukoff2023switchTube} designed a proof-of-concept system with adaptable commitment interfaces for digital well-being.

In contrast, AI-based JITAI leverages large-scale user behavior data and trained AI/ML models to detect appropriate intervention timing and personalize interventions.
For example, Saponaro~\etal~\cite{saponaro2021contextualization} developed two types of AI-based JITAI systems~(population-based, personalized) to reduce users' sedentary behavior. 
Until recently, very few studies explored empowering AI-based JITAI systems with user-in-the-loop to involve user feedback or reactions~\cite{kim_prediction_2022, liao_personalized_2020}.
Mishra~\etal~\cite{mishra2021detecting} implemented an adaptive chatbot that updates the ML model based on users' receptivity to encourage physical activity. Rabbi~\etal~\cite{Rabbi2015} and Liao~\etal~\cite{liao_personalized_2020} integrated reinforcement learning algorithms into JITAI systems to adapt the model to each individual for more effective physical activity intervention. 

To our best knowledge, there have been very few prior studies exploring JITAI-based intervention for smartphone overuse~\cite{lukoff2023switchTube}, not to mention the advanced version that leverages users' reactions in the human-AI loop.
There is a set of technical challenges for such a system. First, the machine learning pipeline needs to respond within seconds when the user opens an app. Second, the model needs to promptly adapt to users' ever-shifting habits.
\projectname{} aims to address these challenges by establishing a real-time human-AI loop system.

\subsection{XAI and Interventions}
\label{sub:background:xai}
Although black-box AI models excel in making complex predictions and handling intricate tasks, they often encounter challenges with interpretability and transparency, making it difficult for users to understand how the models arrive at specific decisions or predictions. This reflects the recent advance of explainable AI (XAI). By providing explanations for AI-driven decisions, XAI not only helps humans comprehend the rationale behind AI system outputs, but also instills a sense of trust and confidence in these systems~\cite{barredo_arrieta_explainable_2020, ribeiro_why_2016,zhang2021effect}. Recent advances in XAI research have not only served AI/ML practitioners and data scientists, enabling them to engage in model debugging and model behavior inspection~\cite{kaur_interpreting_2020, liu2017analyzing}, but have also extended to domain experts in diverse fields~\cite{eiband_bringing_2018} and end-users~\cite{bhatt2020explainable}.
In the context of smartphone overuse detection and intervention, by showing why the users need to stop using certain applications, explanations can help users understand AI decisions and their own device usage patterns.
Moreover, XAI has the potential to activate System 2 within the Dual Process Theory~\cite{hoffman2018metrics}, clarifying the reasoning behind interventions targeting smartphone overuse~\cite{vasconcelos2023explanations}. These explanations stimulate users' deliberate, analytical thinking (System 2) and introduce appropriate reliance on AI~\cite{vasconcelos2023explanations}. This active engagement prompts users to reflect on their usage habits and make conscious adjustments, potentially leading to changes in their smartphone usage behavior.

Despite the considerable attention given to the field of XAI, research at the intersection of XAI and JITAI remains limited.
Woźniak~\etal~\cite{wozniak2020exploring} observed that presenting users with both algorithm-derived fitness goals and a clear explanation for the recommendation could increase their trust towards the recommended goal.
MindScope~\cite{kim_prediction_2022} is a stress management system providing different explanation levels. The results indicated that elaborate explanations helped users understand stress-related events, while categorical explanations allowed them to interpret stressors from their unique perspectives.
In our case, \projectname{} integrates XAI into JITAI to enhance the intervention delivery transparency, effectively address users' confusion about unexpected smartphone interventions, and foster user trust through seamless human-AI interaction.

\section{\projectname{} Design}
\label{sec:design}

We developed \projectname{} --- an intelligent, adaptive, and explainable intervention system for smartphone overuse. Grounded in JITAI principles, the main contribution of our system is the integration of continuous user feedback to form a human-AI loop. The architecture of the \projectname{} system comprises two main building blocks: the ML pipeline to predict smartphone overuse (Section~\ref{sub:design:ml}), and the intricate design of interventions to be shown to users when overuse is predicted (Section~\ref{sub:design:intervention}). 

This section provides an overall introduction to these core system components and the overall intervention flow (Section~\ref{sub:design:flow}).

\subsection{Machine Learning for Smartphone Overuse Prediction}
\label{sub:design:ml}
Constructing an ML-driven JITAI system for predicting smartphone overuse and triggering intervention needs careful design across four key aspects: (1)~feature design, (2)~label collection mechanism, (3)~adaptive model updates, and (4)~explanation generation. In this section, we offer an in-depth exploration of each aspect.

\subsubsection{Feature Design}
\label{subsub:design:ml:feature}
We design a set of five passive sensing feature categories~\cite{meegahapola2023generalization,xu_globem_2022,xu_globem_2022-1} to capture smartphone overuse behavior:

(a) \textbf{Phone and App Usage}. Understanding smartphone overuse requires a thorough analysis of usage patterns. We investigate both the high-level phone usage and the low-level app usage pattern.
For phone usage, we track screen interactions and battery status. Screen interactions encompass phone unlock frequency and duration, calculated from screen-on/off events. As for battery usage, we extract battery consumption rate, charge, and discharge durations. 
For app-related features, our approach includes statistical metrics (count, min, max, mean, standard deviation, sum) linked to app visit frequency and time spent. Additionally, fine-grained user interface interactions (e.g., scrolling, tapping) provide insights into smartphone overuse. We use UI-event-driven features, gathering data on quantities and proportions of events like scrolling, clicking, focusing, and window state changes. Moreover, we also include the count and diversity of notifications.

(b) \textbf{Activity}. Users' interaction with the environment presents another pivotal factor intertwined with smartphone overuse. Concerning attributes related to physical activity, we examine stationary and mobile durations. Furthermore, ambient light offers insights into the user's specific location (such as the room) and may vary depending on the time of day. In the case of ambient light, we extract statistical lux-related features.

(c) \textbf{Social Context}. Users might reduce smartphone interaction in specific social settings, such as when with friends or peers. We primarily focus on text message-derived features (\eg first message time, top contacts) and Bluetooth signals as a proxy of social contexts (\eg mean and standard deviations of scans, unique device counts). While we initially considered call-related features, we excluded them due to limited usage among participants.

(d) \textbf{Location}. Smartphone usage can also be tied to specific places. To capture this, we extract diverse location-based features, including location type, variance, entropy, time at the most-visited places, time at home, and duration of location stays. We also incorporate statistical WiFi data features involving metrics such as scans from visible and connected access points.

(e) \textbf{Time}. Previous work has highlighted different smartphone usage between times (\eg night \vs daytime)~\cite{ahn2014systemic,xu_leveraging_2019}. Therefore, we further include an objective temporal feature to capture this.

\subsubsection{Label Collection Mechanism}
\label{subsub:design:ml:label}
\projectname{} employs supervised learning for smartphone overuse prediction, which needs labeled data to train the model. We gather labels for two purposes: (i) to build the initial ML model and (ii) to update and adapt the ML model over time. Here we present the label collection mechanism for the first step, as shown in Figure~\ref{fig:intervention-interface} (Left). Further information about the data collection approach to update the ML model is provided in Section \ref{subsub:design:ml:adaptive-model} and \ref{sub:design:intervention}.

Our label collection mechanism harnesses Ecological Momentary Assessment~(EMA) methodology~\cite{shiffman2008ecological} by presenting a prompt to ask users to report whether they are overusing their phones. We collect user responses and use them as labels to train our model.
Labels are collected in-the-moment in real-time as users engage with their phones. This instantaneous label collection is crucial as it ensures timely, contextually relevant labels while users' app usage memory and experience remain fresh.

We designed three distinct in-the-moment label collection rules: entry-moment, leaving-moment, and during usage.
Previous studies in smartphone notification management~\cite{ho2005using, iqbal2010oasis, park2017don} showed the effectiveness of breakpoint-based notification delivery. Inspired by these works, we align optimal notification instances with task-switching moments, corresponding to accessing~(i.e., entry-moment) and leaving a monitored app~(i.e., leaving-moment prompts). 
In the smartphone overuse domain, empirical investigations revealed that once involved with a potentially addictive smartphone app, stopping usage becomes challenging, resulting in unforeseen excessive usage~\cite{kim2017technology, ko2015nugu}. 
To address these instances of overuse, we also incorporate label collection during usage, asking users to provide labels every 10 minutes during monitored app usage. Moreover, to alleviate the labeling burden, we implement a cool-down interval. This prevents additional label collection prompts within a cool-down period if a user recently provided a label.
In addition, we also provide a post-hoc labeling process, where users can go back to check any missing annotations of their app usage history.
\footnote{
We recognize that such a label collection mechanism may intervene and affect phone usage behavior. Therefore, in Section~\ref{sec:study}, we intentionally inserted a break week between the label collection and intervention deployment to reduce the impact.
}

\subsubsection{Adaptive Model Updates}
\label{subsub:design:ml:adaptive-model}
We leave the specific ML model choice for the result section (Section~\ref{sec:results}) as it needs empirical evaluation. Here, we focus on the model updates first.
To tailor the ML model to each individual and accommodate their evolving behaviors, dynamic context, and smartphone usage patterns, we regularly update the ML model.
We collect user feedback on the intervention as new labels. Together with the corresponding contextual data, these data can be used to update the ML model.
It is noteworthy that the model update is personalized, \ie we train and update a model for each user.

Updating ML models involves deciding \textit{when} to update the model and \textit{how} to update the model. For the when part, while it's possible to instantly update the model with each new data point, frequent model updates will incur expensive computational costs, causing delays in intervention delivery. We conduct ML updates daily from 12 AM to 1 AM to retain the real-time aspect of just-in-time interventions. This process, taking 2 hours on average, ensures that the adapted models are available by the morning of each intervention day.

How to update the model is another crucial design consideration. A simple approach would involve re-training the model with equal weights for all data samples, treating historical and current user behavior equally. However, this method is sub-optimal as it fails to account for changing user dynamics. In our approach, we adopt decay-based sample weight assignment, \ie recent data will have relatively higher weights.
Specifically, we adopted a linear decreasing assignment from 1.0 to a minimum cap of 0.5. Based on empirical testing with pilot experiments, we assign the weight of the most recent day as 1.0, and it decreases linearly every half-week until it reaches 0.5.
This can help the re-trained model adapt to evolving conditions and smartphone usage behaviors. By emphasizing recent observations, the model becomes more adaptive, effectively capturing current trends while gradually reducing the impact of outdated information.
We discuss other model update methods, such as reinforcement learning, as future work in Section~\ref{sec:discussion}.

\subsubsection{Model Explanation}
\label{subsub:design:ml:model-explanation}
We provide explanations derived from model predictions to enhance users' understanding of the ML-based intervention system's decisions and foster trust and collaboration with AI~\cite{barredo_arrieta_explainable_2020, ribeiro_why_2016}. These explanations are generated based on the top features contributing to an ``overuse'' prediction.
We designed two explanation detail levels: high and low.
A high-level explanation represents the feature category.
As for the low-level explanation, a straightforward option is to use the actual feature name. However, our internal testing found that it introduced unnecessary details and cognitive load. Therefore, we simplify and abstract the raw feature name into a feature description (see Appendix~\ref{sec:appendix:explanation}).
For example, consider the location feature ``time spent at the second most frequent location''. The high-level explanation is ``location'', and the low-level explanation is ``time at frequent locations''.
By default, users will see the high-level explanations and can access more detailed, low-level explanations if they are interested.

\subsection{Intervention Design}
\label{sub:design:intervention}
The JITAI-based intervention system aims to provide \textit{accurate and timely support} while \textit{accommodating shifts in user context and conditions}, as discussed in Nahum~\etal's work~\cite{nahum2018just}. Following these principles, we develop an intervention mechanism based on a typing task (offering the right support level). These interventions are triggered by an intelligent ML model detecting instances of ``overuse''~(optimal timing). Concurrently, user feedback is collected to enhance adaptation to individual user conditions and context~(accommodation). This feedback loop subsequently drives updates to the ML model.
Meanwhile, we also provide explanations derived from the model predictions.

\subsubsection{Intervention Mechanism}
\label{subsub:design:intervention:mechanism}
\begin{figure*}[t]
    \centering
    \includegraphics[width=0.8\linewidth]{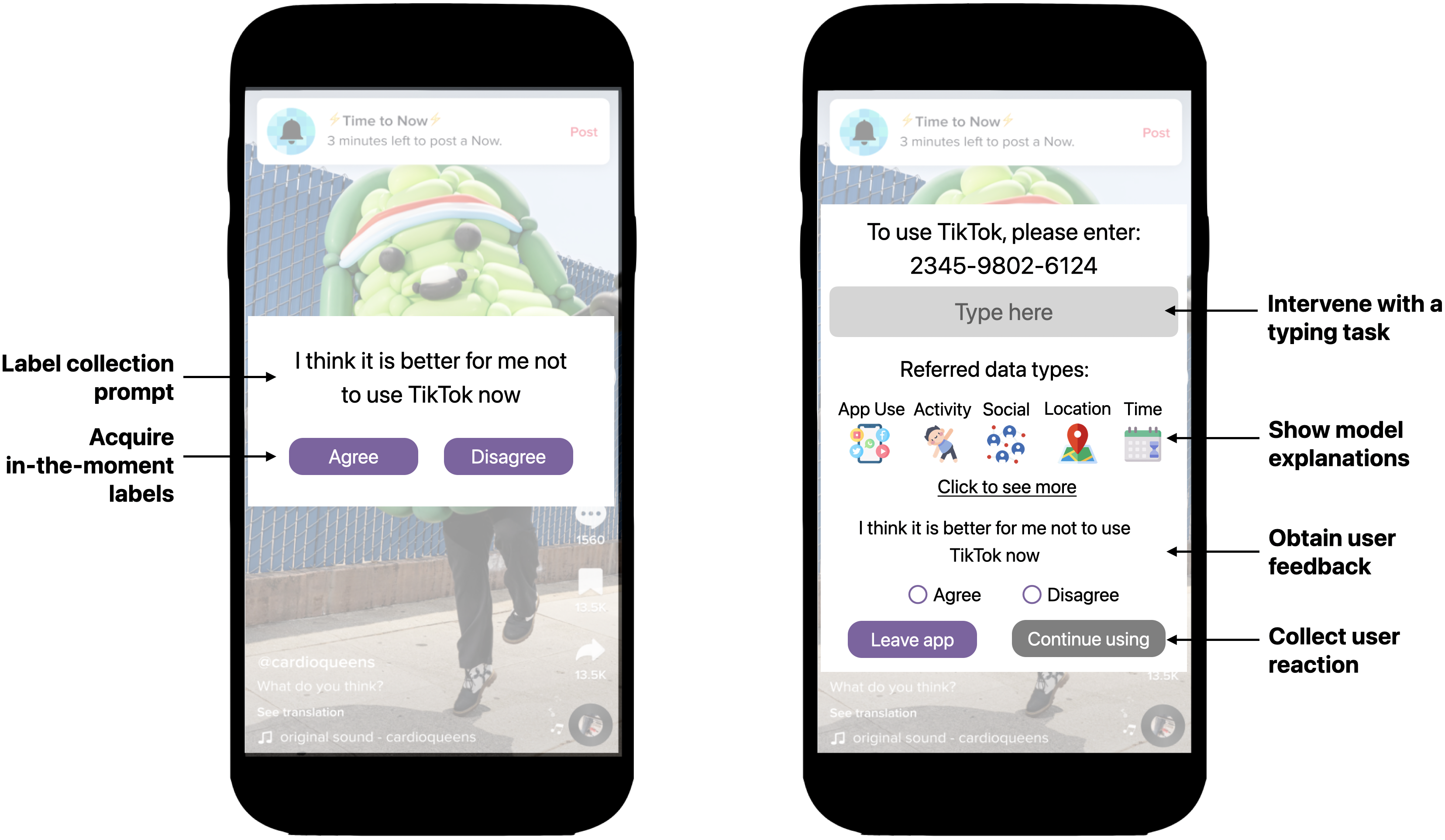}
    \caption{In-the-Moment Labeling and Intervention Interfaces. (Left) In-The-Moment Label Collection Interface; (Right) \projectname{} Intervention Interface. It encompasses four key components from top to bottom: (1) typing-based intervention task, (2) ML model explanations highlighting feature categories aligned with the model's output, (3) collection of user feedback -- this is an optional question that users can choose to respond or ignore, and (4) user actions.}
    \label{fig:intervention-interface}
\end{figure*}

The majority of prior work in the smartphone intervention domain provides interventions by either sending notifications/reminders~\cite{hiniker2016mytime, ko2015nugu} or employing app access restrictions~\cite{kim2019lockntype, ko2016lock}. However, notification-based interventions can be easily circumvented, while excessive restriction may agitate users and lead to counterproductive outcomes. We followed the previous work to balance intervention efficacy and usability and leveraged a typing-based intervention mechanism~\cite{kim2019lockntype, xu_typeout_2022} as interaction friction. 
Users are asked to input specific digits before they can proceed to use a monitored app. The digits are randomly generated within each intervention instance. Users can exit the application anytime and return to the home screen.
Prior work~\cite{kim2019lockntype} suggested that a typing task with a medium workload (10 - 20 digits) was effective and usable. Considering this, we designed a typing task comprising 12 digits.
Note that the specific intervention mechanism is not the main focus of our paper, and we envision our adaptive and explainable system can be integrated with other mechanisms easily.

\subsubsection{Intervention Timing}
\label{subsub:design:intervention:timing}
Our ML model decides whether to intervene based on users' current context and app usage behavior.
Users first select the apps for which they want to receive the intervention (\ie monitored apps), and the intervention will only focus on these apps.
When the model predicts ``overuse'', an intervention interface appears, as shown in Figure~\ref{fig:intervention-interface} (Right).
Moreover, another design choice before triggering intervention involves determining the frequency of feature extraction and model prediction.

Previous Just-in-Time~(JIT)-based smartphone overuse techniques triggered interventions when users launched a monitored app~\cite{kim2019lockntype, lochtefeld2013appdetox}, or when the duration of a target app usage reached a predefined threshold~\cite{hiniker2016mytime, orzikulova2023finerme}.
Our design takes both the launching moment and the usage period into account. We opt to initiate the feature extraction and model prediction process both upon app launch and periodically while the target app is in use. 
We empirically set the prediction interval as 5 minutes based on our pilot study. We further defined a 10-minute cool-down period after triggering an intervention to avoid a disrupted user experience.

\subsubsection{User Feedback to Update Model}
\label{subsub:design:intervention:feedback}
To adapt to dynamic shifts in user context and app usage behavior, we update each individual's ML model regularly. This entails obtaining fresh labels during the intervention period.
One straightforward approach is employing the same label collection mechanism for constructing the initial model (see Section~\ref{subsub:design:ml:label}). However, this would considerably hinder system usability. Users would have to contend with both labeling prompts and intervention notifications. We integrated user labeling within the intervention interface~(Figure~\ref{fig:intervention-interface} Right) to address this issue.

When the intervention pops up, users are encouraged to provide feedback with a simple click to indicate whether they are overusing the phone. 
We design the labeling prompt with simplicity while ensuring it provides guidance to identify instances of smartphone overuse.
The phrasing is deliberately structured to avoid potentially eliciting negative feelings regarding users' behavior. Rather than posing a direct query about smartphone or app overuse, users are prompted to indicate their agreement or disagreement with the statement: ``I think I shouldn't use \texttt{AppName} now.'' In cases of agreement, the data point is categorized as ``overuse'', which can be used to reinforce the ML model; conversely, in instances of disagreement, it is classified as ``not overuse'', which can serve as a correction to the model.
Once we receive feedback, we utilize them as new labels to update the ML model, following the design we introduced in Section~\ref{subsub:design:ml:adaptive-model}.
Note that this is an optional question, and users are not forced to respond.
This process can capture the false positive cases, \ie an intervention pops up when users are not overusing their phones. 
Moreover, users can also leverage the post-hoc labeling to provide feedback on false negative cases, \ie an intervention does not pop up when they are overusing their phones.

\subsubsection{Model Explanations}
\label{subsub:design:intervention:explanation}
As detailed in Section~\ref{subsub:design:ml:model-explanation}, our explanation framework generates explanations at two levels: high and low. Previous XAI-based JITAI work in stress management by Kim~\etal~\cite{kim_prediction_2022} showed that although most users favored more detailed explanations, such low-level explanations could potentially undermine the system's trustworthiness. Based on these findings, we decided to highlight the categories of essential features, such as ``location'', ``activity'', ``app usage''~(see Figure~\ref{fig:intervention-interface} Right).
The interface only presents the top three crucial feature categories for the ML model inference and hides other categories to avoid confusion. We use their high-level explanations as icons in Figure~\ref{fig:intervention-interface}.
Furthermore, we provide low-level feature descriptions for users seeking deeper insights by clicking the ``Click to see more'' button.

\subsection{Intervention Flow}
\label{sub:design:flow}
Combining the ML and intervention design in Section~\ref{sub:design:ml} and \ref{sub:design:intervention}, the intervention flow of \projectname{} is visualized in Figure~\ref{fig:system-overview}. There are two loops within the flow: (1)~the inner loop (green) is dedicated to the ML model inference process, and (2)~the outer loop (blue) manages the ML model update process.

In the inner loop, ML model inference is performed through a sequence of steps.
\textcircled{1}~Contextual and app usage data are initially collected by the mobile app (depicted on the left) and transmitted to the cloud server.
\textcircled{2}~Here, the cloud server pre-processes raw data, extracts features, performs inference, obtains prediction output, and generates corresponding explanations (depicted on the right of Figure~\ref{fig:system-overview}). The output of the model's prediction and explanations are then relayed to the user. In cases where the model predicts ``overuse'', the intervention interface~(as illustrated in Figure~\ref{fig:intervention-interface}) will pop up.

Conversely, the outer loop takes charge of the ML model update through user feedback and model enhancement cycles.
\textcircled{3}~When the interface appears, users can provide feedback indicating whether they are overusing their phones.
\textcircled{4}~This feedback is then transmitted to the cloud server, where new labels and features are re-trained. The updated ML model is then employed to generate more tailored and adaptive predictions, which are conveyed back to the user.

\section{System Implementation}
\label{sec:implementation}
Based on the system design in Section~\ref{sec:design}, we then introduce the implementation details of \projectname{}.
We instantiated \projectname{} on Android OS (end-user side) and a server (cloud side), as shown in Figure \ref{fig:mobile-cloud-communication}.
We conducted a one-week pilot field study with four authors of this paper to debug and finalize the system implementation, which includes the sensing platform (Section~\ref{sub:implementation:context-sensinig}), the intervention interface (Section~\ref{sub:implementation:interface}), and the ML pipeline (Section~\ref{sub:implementation:ml}).

\subsection{Context Sensing}
\label{sub:implementation:context-sensinig}
To obtain the data we mentioned in Section~\ref{subsub:design:ml:feature}, 
we leverage AWARE, an open-source passive sensing platform designed for behavioral data collection~\cite{ferreira2015aware}. Our data collection includes multiple sensor streams: location, Bluetooth, Wi-Fi, network, light, screen activity, activity recognition, and communication (including SMS and calls).
We further build our custom app usage tracker with Android's \texttt{AccessibilityService} API~\cite{accessibilityAPI} that adeptly identifies the start and end of app sessions, dynamically monitors time allocation and visit frequencies, captures notifications from monitored applications, and records fine-grained user interactions with monitored apps, including scrolling, clicking, focusing, and window state changes.

\subsection{Intervention Interface}
\label{sub:implementation:interface}
We have introduced the interface design in Section~\ref{sub:design:intervention}, making the Android implementation straightforward.
Moreover, the interface is implemented as an \texttt{AlertDialogue}, which becomes an overlay on top of the monitored app.
When users enter the displayed random digits correctly into the input form and click the ``Continue using'' button, the overlay window is dismissed, and users are allowed to use the app. Conversely, upon clicking the ``Leave app'' button, the phone programmatically returns to the home screen.
Users' reactions to the intervention will not impact the content in the app.

\subsection{Machine Learning Pipeline}
\label{sub:implementation:ml}
\begin{figure*}[t]
    \centering
    \includegraphics[width=0.8\linewidth]{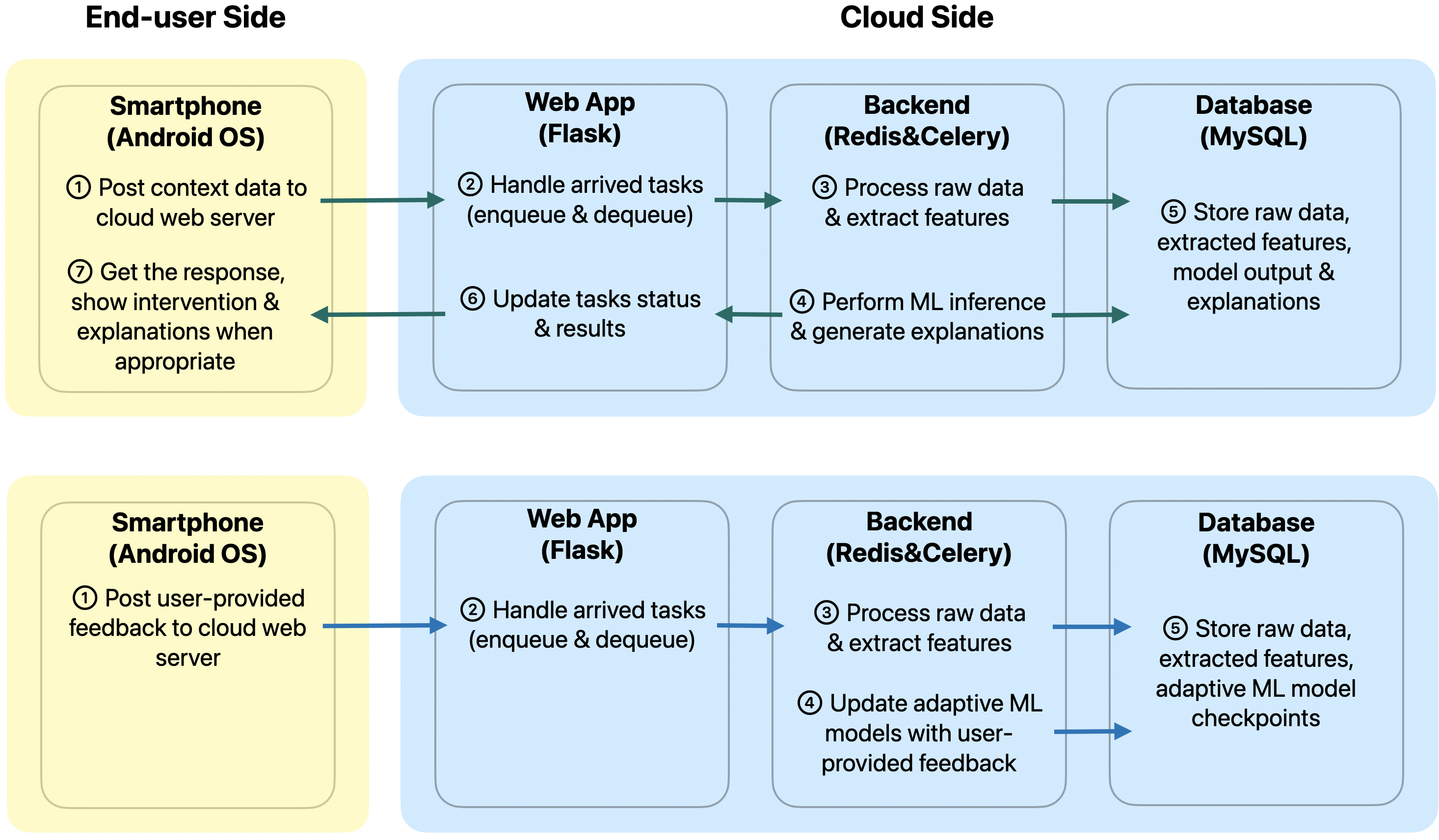}
    \caption{Overview of System Implementation. (Top): Model Inference. (Bottom): Model Update Leveraging User Feedback.}
    \label{fig:mobile-cloud-communication}
\end{figure*}
Our ML pipeline consists of three parts: (1) model inference, (2) model update, and (3) explanation generation. The technical details of these components are described in Figure~\ref{fig:mobile-cloud-communication}.  
The end-user side is a mobile app running on Android OS, and the cloud side consists of a web app (Flask), a back-end (Redis and Celery), and a database (MySQL). The upper figure describes the model inference, and the lower sub-figure describes the model update. 

\subsubsection{Model Inference}
\label{subsub:implementation:ml:inference}
The pipeline contains seven steps. 
\textcircled{1}~The Android client posts the contextual data to the cloud web server with the frequency described in Section~\ref{subsub:design:intervention:timing}.
\textcircled{2}~A Flask-based web app manages a task queue that handles the arrived tasks. Once the task arrives at the cloud, the web app enqueues the task and dequeues in a first-in-first-out (FIFO) manner. \textcircled{3}~The back-end processes the raw data by imputing missing values, normalizing the raw values, and extracting features.
\textcircled{4}~It also performs the inference using the ML model to obtain overuse prediction. Explanations are generated using the SHAP method (see Section~\ref{subsub:implementation:ml:explanation})~\cite{lundberg_unified_2017}.
\textcircled{5}~Raw data, extracted features, model outputs and generated explanations are then stored in the cloud database.
\textcircled{6}~Once model outputs and explanations are ready, the web app updates the task status and the results so that the client can pick it up.
\textcircled{7}~The Android client sends a request to obtain the results. In our pilot study, most of the responses arrived within 3 seconds. If the user is still using the monitored app when the results arrive, it checks the model output.
If the prediction is ``overuse'', the Android client will pop up the intervention, together with the feature explanations. Otherwise, no intervention will show up.

\subsubsection{Model Update}
\label{subsub:implementation:ml:update}
\begin{sloppypar}
As introduced in Section~\ref{subsub:design:ml:adaptive-model} and \ref{subsub:design:intervention:feedback}, \projectname{} updates the model on a daily basis. This pipeline includes five steps. \textcircled{1} User-provided feedback is stored in the mobile app and sent to the cloud server.
The next two steps of task handling (\textcircled{2}) and feature extraction (\textcircled{3}) are similar to the ones in model inference. \textcircled{4}~Next, the adaptive ML model is re-trained, using the user-provided feedback as new labels. We adopt the weight assignment introduced in Section~\ref{subsub:design:ml:adaptive-model} during the re-training.
\textcircled{5}~Lastly, all data, extracted features, and the new model checkpoint are stored in the database.  
\end{sloppypar}

\subsubsection{Explanation Generation}
\label{subsub:implementation:ml:explanation}
To interpret the model predictions, we measure feature importance with SHapley Additive exPlanations (SHAP)~\cite{lundberg_unified_2017}, an XAI method that computes the impact of each feature on prediction outcomes.
We rank the features based on their importance and obtain the corresponding high-level and medium-level explanations introduced in Section~\ref{subsub:design:ml:model-explanation} and \ref{subsub:design:intervention:explanation}.
These explanations are sent to the user along with the model outcome during the model inference.

\section{Field Experiment}
\label{sec:study}
To investigate how AI-powered intelligent and explainable JITAI can affect smartphone overuse in real-life scenarios, we conducted an 8-week field experiment using \projectname{}.
Our study aims to evaluate both the adaptive aspect and explainable aspect of \projectname{}, which requires careful experiment design (Section~\ref{sub:study:design}). We then introduce our field experiment procedure (Section~\ref{sub:study:procedure}) and participants (Section~\ref{sub:study:participants}).

\begin{table}[t]
\caption{Multiple Intervention Types with Characteristics. The last row represents our complete \projectname{} system with ML-powered adaptive and explainable JITAI.}
\label{tab:intervention-type-description}
\small 
\begin{tabular}{c|c|c|c}
\hline \hline
\multirow{2}{*}{\textbf{Intervention Type}} & \multicolumn{3}{c}{\textbf{Characteristics}} \\ 
\cline{2-4} 
& \textbf{ML-based} & \textbf{Adaptive} & \textbf{Explainable} \\ 
\hline
Control & \XSolidBrush & \XSolidBrush & \XSolidBrush \\
\hline
Personalized & \Checkmark & \XSolidBrush & \XSolidBrush \\
\hline
Adaptive-wo-Exp & \Checkmark & \Checkmark & \XSolidBrush \\
\hline
\begin{tabular}[c]{@{}c@{}}Adaptive-w-Exp \\ (\ie \projectname{})\end{tabular} & \Checkmark & \Checkmark & \Checkmark \\    
\hline \hline
\end{tabular}
\end{table}

\subsection{Experiment Design}
\label{sub:study:design}

\subsubsection{How to Evaluate Adaptive and Explainable Interventions?}
To assess the efficacy of the adaptive and explainable components, we devised four distinct intervention types, each taking one step more advanced than the previous method (see Table~\ref{tab:intervention-type-description}).

(1) \textbf{\textit{Control}}. This was a baseline method. It intervened with users simply based on probability (\eg a user might receive intervention when launching an app and every five minutes in 30\% of the cases). The individual probability of the intervention was derived from the user-provided labels during the first phase of the experiment (the modeling phase, see Section~\ref{sub:study:procedure}).

(2) \textbf{\textit{Personalized}}. This method added the ML component on top of \textit{Control}, using the data collected during the modeling phase.
To ensure a \textit{Personalized} model aligned with each user's behavioral patterns while leveraging the rich data from other users, greater emphasis was placed on the user's own data by assigning it higher weights than the data collected from others.
Through empirical tests, the weight for self-data was set at 1.0, while others' data received a weight of 0.1. The personalized model remained static and unchanged throughout the intervention period.

(3) \textbf{\textit{Adaptive-wo-Exp}}. This method further added the adapting component on top of \textit{Personalized}. The model underwent a similar training procedure as \textit{Personalized} at first. It also involved daily model re-training and updates, using continuous user feedback and corrections in response to intervention prompts, as we introduced in Section~\ref{subsub:design:ml:adaptive-model}.

(4) \textbf{\textit{Adaptive-w-Exp}}. Finally, this method added the explanation component on top of \textit{Adaptive-wo-Exp} and completed the whole \projectname{} system. The model of \textit{Adaptive-w-Exp} was identical to that of \textit{Adaptive-wo-Exp}. The only distinction was that \textit{Adaptive-w-Exp} provided ML output explanations in the intervention interface, as introduced in Section~\ref{subsub:design:ml:model-explanation} and shown in Figure~\ref{fig:intervention-interface}.

Note that both \textit{Adaptive-wo-Exp} and \textit{Adaptive-w-Exp} fell into the category of adaptive models. Other than \textit{Adaptive-w-Exp} that displayed explanations, the interface of the other three types was exactly the same to reduce bias.

\subsubsection{Micro-Randomized Trials} 
Considering the sample size to compare four groups, we adopted a within-subject design.
Specifically, we employed Micro-Randomized Trials, which is an experimental design technique optimized for JITAI-grounded intervention within the mHealth domain~\cite{klasnja2015microrandomized}. Instead of having users go through different experiment groups one by one, this method proposes to randomize the groups with smaller units (\eg daily or each intervention), so that the effect of potential confounding variables can be reduced.

In our case, we altered the intervention type among the four types on a daily basis, and each participant experienced only one type of intervention every day. In order to minimize the order effect, we employed the Latin Square design ($n$=4)~\cite{edwards1951balanced} to diversify the intervention altering order.
During our study onboarding sessions, we briefly introduced the four intervention types to users, but they were not informed of the specific order or dates for the four intervention types during the field study.  This was also designed to reduce cognitive bias.

\subsubsection{Evaluation metrics}
We focused on four quantitative metrics to evaluate the performance of \projectname{} and the other three intervention types. The first two were about the model performance: (1) intervention accuracy, (2) intervention receptivity. The other two focused on its impact on users' phone usage patterns: (3) app usage duration and (4) app visit frequency.

Specifically, intervention accuracy represented the proportion of interventions that were marked as ``correct'' by users among the total number of intervention pop-ups. Note that this is a subjective algorithm measure instead of an objective measure, as there is no way to obtain an objective ground truth of overuse.
Intervention receptivity, on the other hand, referred to users' reaction after encountering interventions, which included stopping usage (\eg returning to the home screen, triggering a screen-off event by locking the phone) or continuing usage. Instances where users quit the app were considered receptive interventions, while instances of continued usage were designated as non-receptive interventions.
The other two metrics of app usage duration and visit frequency were calculated from the collected app usage log.

For qualitative metrics, we revealed the exact dates for each intervention type to users at the end of the study. We highlighted the latest four days to help users recall their experience with the four different techniques, as they had the most fresh memory. Then, we distributed a final questionnaire, asking them to rank the four types based on their preferences, as well as their perceived accuracy, effectiveness, and level of trust in different intervention types.
Moreover, we conducted semi-structured exit interviews with participants to collect their feedback and intervention preferences.
Our interview started with questions: ``What do you think of the four intervention techniques? What's the reason behind your preference ranking? What do you think of the explanations coming with interventions?'' For participants with low intervention accuracy and receptivity, we also asked about their thoughts and reactions towards intervention. We then followed the participants' lead and followed up with more detailed questions.
The interviews were recorded, and three researchers followed the procedure of thematic analysis~\cite{braun2012thematic} to independently analyze and code the data. Then, they met, discussed, and iterated the coding until convergence.

\subsection{Procedure}
\label{sub:study:procedure}
\begin{figure*}[t]
    \centering
    \includegraphics[width=0.75\linewidth]{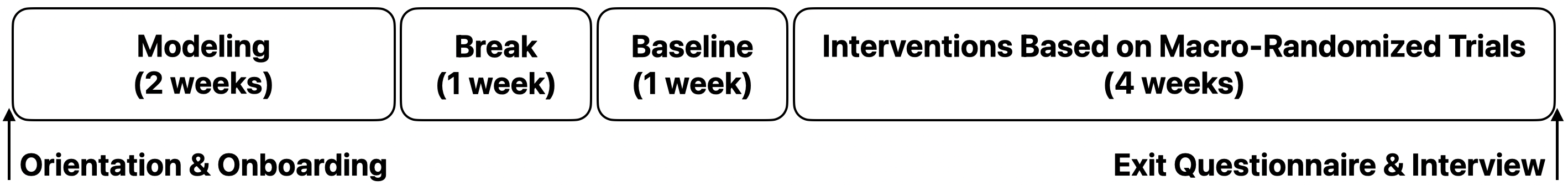}
    \caption{Field Experiment Flowchart}
    \label{fig:study-flowchart}
\end{figure*}

Our field experiment consisted of eight weeks, as shown in Figure~\ref{fig:study-flowchart}.
After the orientation and onboarding session, our field deployment experiment followed a sequence of four phases: (1) an initial modeling phase involving label collection lasting for two weeks, (2) a one-week break phase, (3) a subsequent week dedicated to baseline data collection without any intervention, and (4) a final four-week intervention phase with the design of micro-randomized trials.

In the modeling phase, we passively collected contextual and app usage data along with user-provided labels, using the label collection mechanism (see Section~\ref{subsub:design:ml:label}). These data points were used to calculate the individual probability (for \textit{Control}) and train the initial ML models (for \textit{Personalized}, \textit{Adaptive-wo-Exp}, and \textit{Adaptive-w-Exp}).
To mitigate carry-over effects inherent in label collection, we incorporated a designated break phase.
Then, we proceeded with the baseline week, during which we gathered baseline app usage data (usage duration and visit frequency of monitored apps) when there was no intervention. This data would serve as a comparative benchmark against diverse intervention types.
Finally, during the intervention stage, interventions were introduced to users. User feedback and the accompanying behavioral data were collected (see Section~\ref{subsub:design:intervention:feedback}) to update the \textit{Adaptive-wo-Exp} and \textit{Adaptive-w-Exp} models. Since users could still provide feedback during the \textit{Control} and \textit{Personalized}, these data were also collected to update the adaptive models.

At the end of the intervention phase, the intervention order was presented to participants. They then filled out the questionnaire and completed the exit interview. They were compensated up to \$50 based on their study compliance.

\subsection{Participants}
\label{sub:study:participants}
We posted a call for participation on large university community forums, together with a survey including basic demographics and a Smartphone Addiction Scale (SAS, score ranging from 33 to 198)~\cite{kwon2013development}.
We selected participants who used an Android smartphone as their primary phone and had a high SAS score (>120).
176 participants met the criteria. 127 of them attended the onboarding session.
Among these participants, 49 discontinued their participation during the field study. Out of the 49 discontinued participants, 20 chose to exit the study citing personal reasons, 17 encountered software and hardware issues, 8 experienced compatibility concerns, 3 raised privacy issues, and 1 attributed their departure to battery concerns.
Seven participants whose sensor or usage data only covered three or fewer intervention types were also eliminated from the analysis.
In total, 71 participants (48 females, and 23 males, aged 21.8 $\pm$ 2.3, from 18 to 27) completed the whole study and provided high-quality data. Our analysis results were based on these participants.

\section{Results}
\label{sec:results}
Throughout our field experiment, we collected 497,458 minutes of usage data for 149 monitored apps (17 $\pm$ 5 apps per person) from 207,898 app sessions.
App categories of entertainment, social media, and shopping emerged as the most frequently selected app categories.
In total, we collected 75,670 ground truth labels during the modeling phase. 60.5\%, 24.5\%, and 14.9\% of them were collected at the entry, using, and exit stages.
During the intervention phase, we captured 47,939 intervention encounters, among which we collected 39,188 additional labels from user feedback.
These data were used for our quantitative analysis. We also investigated the qualitative data from questionnaires and interviews.

To build the optimal initial AI-based intervention models, we first compared multiple ML models using the data from the modeling phase (Section~\ref{sub:results:ml-comparison}).
After checking the intervention frequency among different intervention methods (Section~\ref{sub:results:intervention_frequency}), we then evaluated the adaptiveness and explanation aspects of \projectname{} from multiple metrics, including accuracy and receptivity (Section~\ref{sub:results:intervention}), app usage duration and visit frequency (Section~\ref{sub:results:usage}), as well as participants' perceived effectiveness of different intervention types (Section~\ref{sub:results:subjective}).
Overall, our findings showed the consistent advantage of the adaptive component (\textit{Adaptive-w-Exp}/\textit{Adaptive-wo-Exp} \vs \textit{Personalized}/\textit{Control}). We also observed interesting effects of explanations (\textit{Adaptive-wo-Exp} \vs \textit{Adaptive-w-Exp}) on app usage behavior and user experience.

\subsection{ML Model Comparison}
\label{sub:results:ml-comparison}

Using the data gathered during the modeling phase, we compared a wide range of off-the-shelf ML models, including Naive Bayes (NB), Logistic Regression (LR), Support Vector Machines (SVM), Decision Trees (DT), Random Forest (RF), and K-Nearest Neighbors (KNN).
To account for real-world temporal changes in user behavior and simulate actual model deployment, we used the first week for training and the subsequent week for testing.

The collected data was imbalanced (37.8\% overuse, 62.2\% non-overuse). Other than calculating individual probabilities for the \textit{Control} intervention type (42.4 $\pm$ 24.4\%), we experimented with SMOTE-based under-sampling and up-sampling methods for model training~\cite{chawla2002smote}.
We also tuned hyperparameters on promising models with grid search.
Our results indicated that RF (number of estimators: 100, max depth: 10, min samples split: 5), coupled with the up-sampling method, had the best performance across all models, with an F1 score of 66.7\%. Other models had worse results: NB (55.3\%), LR (59.0\%), SVM (59.0\%), DT (59.6\%), KNN (62.6\%). 
We use this RF model as the static ML model for \textit{Personalized}, as well as the initial model for \textit{Adaptive-wo-Exp} and \textit{Adaptive-w-Exp}.

We also performed a feature importance analysis across all users' models. Our analysis revealed consistency in vital features among participants: the most common important features were related to phone usage (unlock duration), location  (total travel distance, moving to static ratio), and temporal feature (\eg whether night time).

\subsection{Intervention Frequency}
\label{sub:results:intervention_frequency}
Prior to the comparison of intervention effectiveness, we first compare the frequency of intervention in our field experiment.
Our Friedman test across four intervention types showed that the number of daily interventions was significantly different ($\chi^2 = 16.60$, $p < 0.001$). Our post-hoc pairwise comparison (Wilcoxon signed-rank test with Holm-Bonferroni correction) indicated differences between \textit{Control} and all the rest threes ($p$s $<0.01$), but not for other pairs. This means that the personalized and adaptive models sent fewer interventions to users.
As shown in the rest of this section, they were more effective with less intervention frequency.

\subsection{Intervention Accuracy and Receptivity}
\label{sub:results:intervention}
In this section, we investigate the effectiveness of adaption (Section~\ref{subsub:results:intervention:adapation}) and explanation (Section~\ref{subsub:results:intervention:explanation}) through the perspective of intervention accuracy and receptivity. We also measure the performance dynamics over time (Section~\ref{subsub:results:intervention:receptivity-time}).
Since individual behaviors varied greatly across participants, we used \textit{Control} as the benchmark and normalized accuracy and receptivity metrics for each participant accordingly. 
A value higher than 1.0 means better performance and a value lower than 1.0 indicates worse performance.

\subsubsection{Effectiveness of Adaptation}
\label{subsub:results:intervention:adapation}

\begin{figure}[t]
    \centering
    \begin{minipage}[b]{0.4\textwidth}
        \centering
        \includegraphics[width=\textwidth]{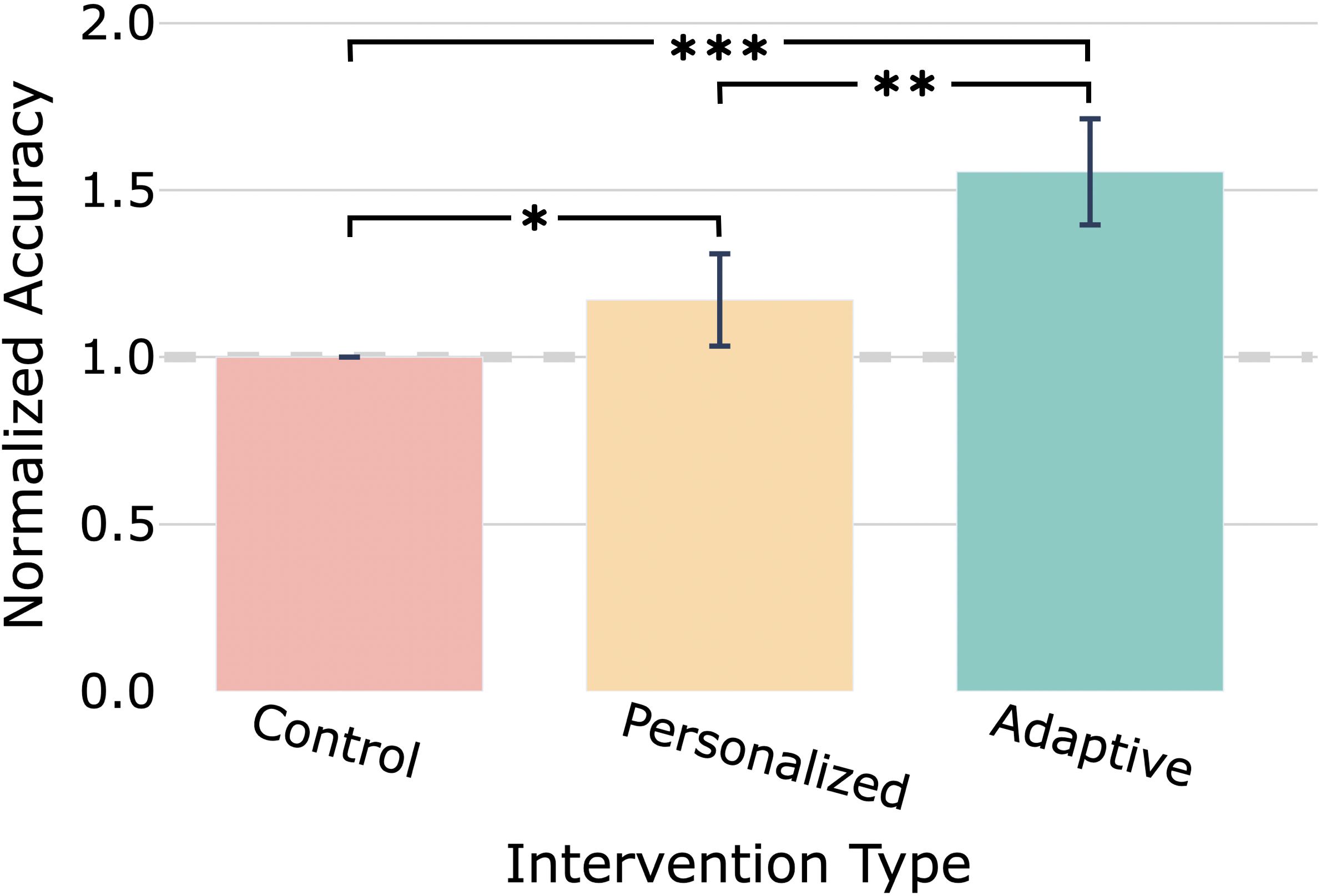}
    \end{minipage}
    \hspace{0.3cm}
    \begin{minipage}[b]{0.4\textwidth}
        \centering
        \includegraphics[width=\textwidth]{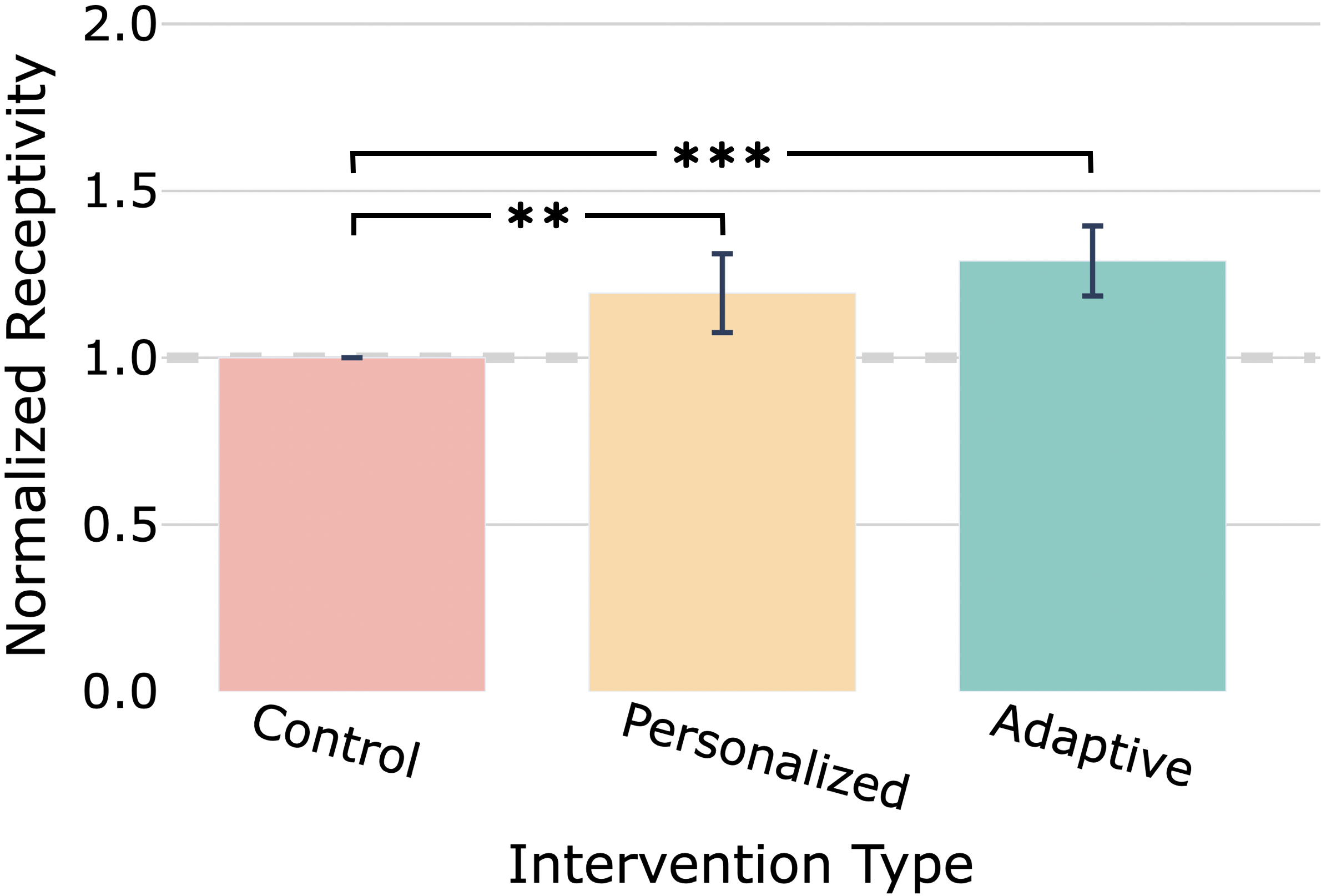}
    \end{minipage}
    \caption{Intervention Accuracy (Top) and Receptivity (Bottom) Comparison across Three Intervention Types. Error bar indicates standard deviation. The same below.
    The two adaptive versions (with and without explanation) are merged into \texttt{Adaptive} to highlight better that adaptive ML-based methods had higher intervention accuracy and receptivity.}
    \label{fig-intv-acc-and-rec-agg-3types}
\end{figure}

\textbf{Our results indicated that adaptive methods achieved significantly higher intervention accuracy and receptivity.}
To evaluate the effectiveness of our intelligent intervention types (static or adaptive ML models), we first merged \textit{Adaptive-wo-Exp} and \textit{Adaptive-w-Exp} into a type called \textit{Adaptive} to highlight the adaptation property better.
Figure~\ref{fig-intv-acc-and-rec-agg-3types} (Left) shows the comparison across the three types \textit{Control}, \textit{Personalized} ($\Delta$=17.1\% over \textit{Control}), and \textit{Adaptive} ($\Delta$=55.5\%).
We fitted a Generalized Linear Mixed Model~(GLMM) on intervention accuracy, with the Gamma family based on a Kolmogorov–Smirnov distribution test\footnote{Unless noted otherwise, we repeated the same procedure for the rest of the GLMM models.}. We set intervention type as the main effect and participant ID as the random effect.
Our results showed that the intervention type had a significant effect ($\chi^2(2)$=24.52, $p$<0.001). Post-hoc analysis with Holm-Bonferroni correction further indicated that both the static \textit{Personalized} model~($p$<0.05) and the \textit{Adaptive} models~($p$<0.001) had significantly higher intervention accuracy compared to the \textit{Control} baseline.
Furthermore, the \textit{Adaptive} model further significantly outperformed the static \textit{Personalized} model~($p$<0.01, $\Delta$=32.8\%).
These results not only revealed the effectiveness of the ML component (\textit{Personalized} \vs \textit{Control}), but also more importantly, indicated the effectiveness of the adaptation part (\textit{Adaptive} \vs \textit{Personalized}).

While accuracy refers to explicit user subject feedback on interventions, receptivity describes their actual behavior~(\ie continue using the app or quitting it).
Hence, receptivity metrics enable us to measure how different interventions affect participants' actual behavior.
We ran another GLMM on the intervention receptivity with the same setup as the accuracy test. Similarly, the results also indicate the significance of intervention type on receptivity ($\chi^2(2)$=18.44, $p$<0.001), as shown in Figure~\ref{fig-intv-acc-and-rec-agg-3types} (Right).
The post-hoc\break  pairwise results indicated that participants were more receptive when using the \textit{Personalized} ($p$=0.005, $\Delta$=19.4\%) and \textit{Adaptive} ($p$<0.001, $\Delta$=29.0\%) intervention types compared to the \textit{Control}.
These observations on the receptivity metric were consistent with those in the accuracy metric.

\subsubsection{Effectiveness of Explanations}
\label{subsub:results:intervention:explanation}
\begin{figure}[t]
    \centering
    \begin{minipage}[b]{0.4\textwidth}
        \centering
        \includegraphics[width=\textwidth]{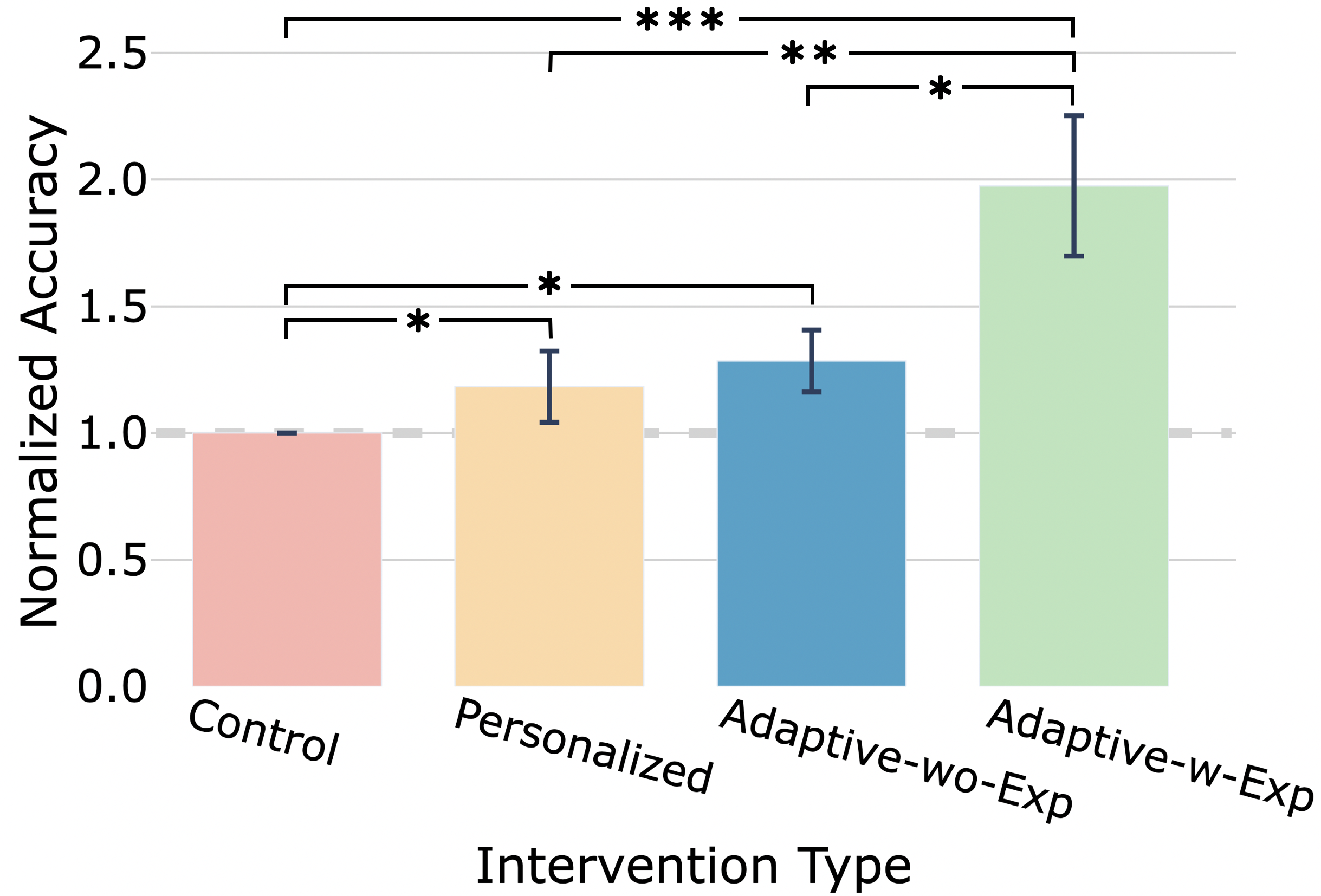}
    \end{minipage}
    \hspace{0.2cm}
    \begin{minipage}[b]{0.4\textwidth}
        \centering
        \includegraphics[width=\textwidth]{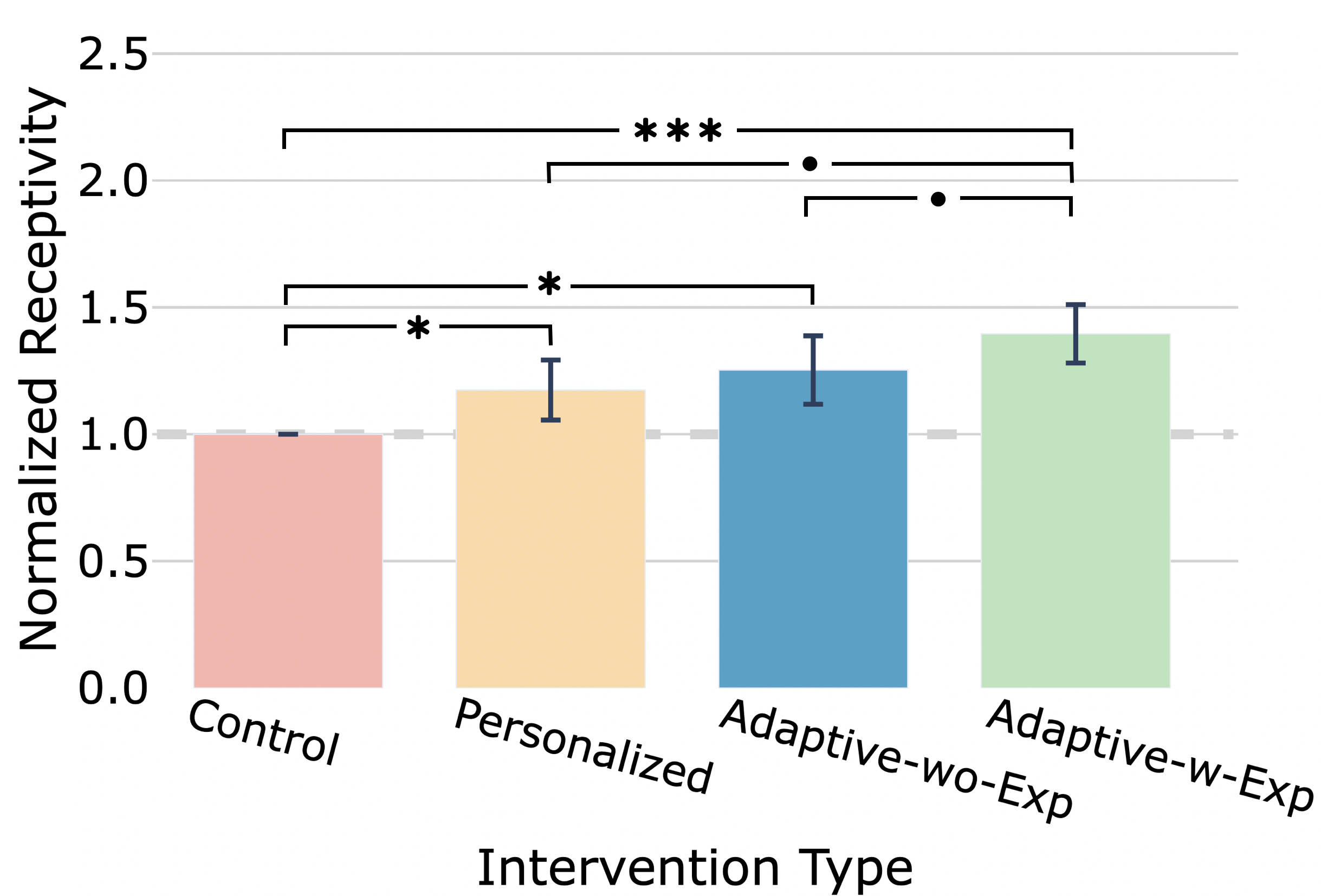}
    \end{minipage}
    \caption{Intervention Accuracy (Top) and Receptivity (Bottom) Comparison across Four Intervention Types. The two versions of \texttt{Adaptive} are divided (\textit{Adaptive-w-Exp} and \textit{Adaptive-wo-Exp}) to better highlight that adding explanations can further enhance the performance of interventions.}
    \label{fig:intv-acc-and-rec-agg-4types}
\end{figure}

\textbf{Our results suggested that adding explanations significantly enhanced intervention accuracy and receptivity.}
To investigate the impact of explanations, we divided the \textit{Adaptive} type back to the original two groups \textit{Adaptive-wo-Exp} and \textit{Adaptive-w-Exp}.
The comparison results of the four intervention types are shown in Figure~\ref{fig:intv-acc-and-rec-agg-4types} (Left).
We ran another GLMM on accuracy, with the four intervention types as the main effect and participant ID as the random effect.
The results showed significance of intervention types ($\chi^2(3)$=35.70, $p$<0.001), and the post-hoc analysis suggested that \textit{Adaptive-w-Exp} (\ie our complete \projectname{} system) interventions exhibited the highest accuracy by outperforming \textit{Control} ($p$<0.001, $\Delta$=97.5\%), \textit{Personalized} ($p$<0.01, $\Delta$=66.9\%), and even \textit{Adaptive-wo-Exp} ($p$<0.05, $\Delta$=53.8\%).
This evidence suggested the effectiveness of explanations: By explaining why they might be overusing smartphones, \projectname{} could help participants better realize and recognize their overuse behavior than the cases without explanations.

Similar to accuracy, our GLMM on receptivity also showed significance ($\chi^2(3)$=25.57, $p$<0.001). \textit{Adaptive-w-Exp} also achieved the highest receptivity, as shown in Figure~\ref{fig:intv-acc-and-rec-agg-4types} (Right), with strong significance over \textit{Control} ($p$<0.001, $\Delta$=39.6\%), as well as marginal significance over \textit{Personalized} ($p$=0.06, $\Delta$=18.9\%) and \textit{Adaptive-wo-Exp} ($p$=0.07, $\Delta$=11.4\%).
Combining the results of both intervention accuracy and receptivity, we found that \projectname{} could not only help participants recognize their overuse behavior (higher accuracy), but also help them stop using an app in the moment (higher receptivity). This finding suggests the effectiveness of \projectname{} by delivering interventions when the users were receptive~\cite{nahum-shani_just--time_2018, mishra_detecting_2021}.

\subsubsection{Effectiveness over Time}
\label{subsub:results:intervention:receptivity-time}
\begin{figure*}[b]
    \centering
    \includegraphics[width=0.9\linewidth]{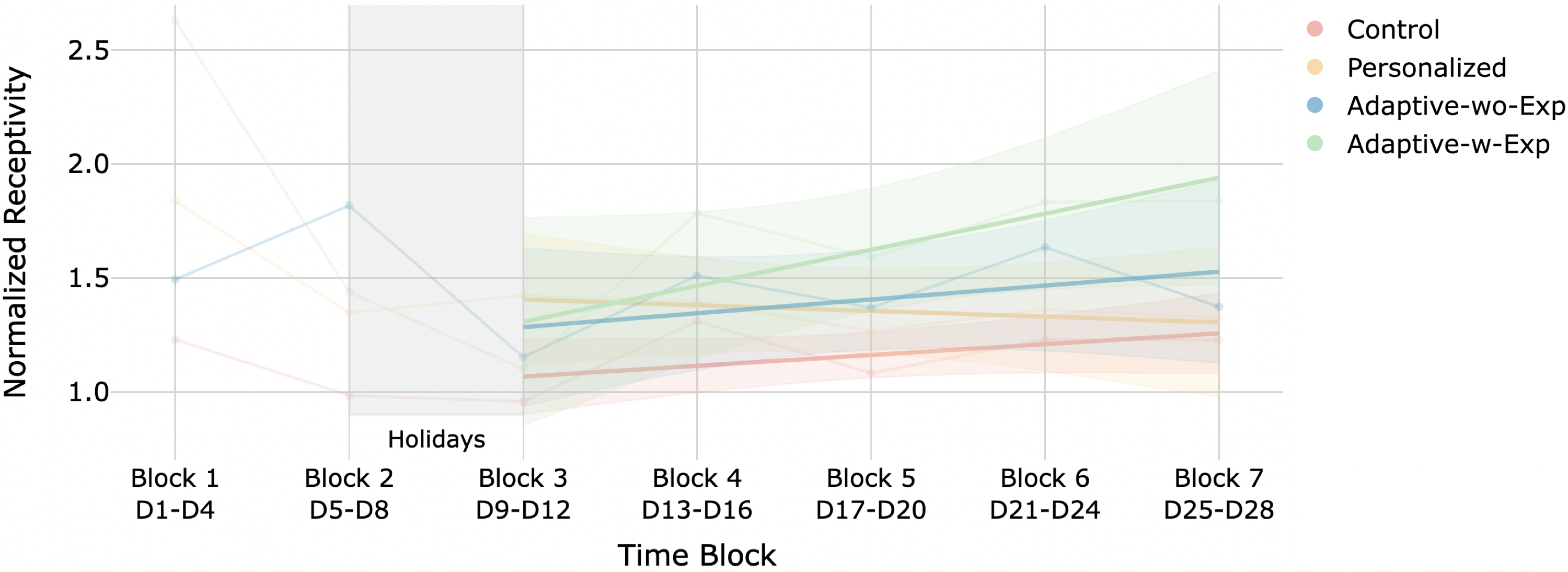}
    \caption{Intervention Performance Over Time. Both adaptive models had an increasing trend, followed by the \textit{Control} group, while \textit{Personalized} method showed a decreasing trend.}
    \label{fig:receptivity_overtime}
\end{figure*}
\textbf{Both adaptive models had increasing intervention performance over time.}
\textit{Adaptive-wo-Exp} and \textit{Adaptive-w-Exp} approaches both regularly updated the ML model nightly. We also evaluated their intervention performance as the field study progressed.
As intervention receptivity provides a more objective reflection on user behavior, we analyzed receptivity dynamics over time, as presented in Figure~\ref{fig:receptivity_overtime}.
The Y-axis represents normalized receptivity, while the X-axis denotes the progress of the intervention phase.
Since we used micro-randomized trials, we took four days as an intervention block, constituting a complete cycle of four distinct intervention types.
Block 2 coincided with a national holiday period, during which participants were on a break and did not attend classes.
We observed a significant drop in receptivity from Block 1 to 2 (see the left of Figure~\ref{fig:receptivity_overtime}. Thus, we focused our analysis after Block 3.

We observed an increasing trend in receptivity for the two adaptive intervention types, with \textit{Adaptive-w-Exp} having the most positive slope ($r$=0.16, $\Delta$=63.6\%), followed by \textit{Adaptive-wo-Exp} ($r$=0.06, $\Delta$=19.1\%). These results indicated that our adaptive models could gradually improve over time and that explanations could continuously enhance the intervention's effectiveness.
Moreover, \textit{Personalized}'s receptivity was consistently higher than the \textit{Control} baseline across all blocks. However, \textit{Personalized} showed a slight decreasing trend ($r$=-0.03), while \textit{Control} showed a slight increasing trend ($r$=0.05). This result may indicate that participants got used to the static ML-based intervention and had less receptivity over time. Our interview data revealed the potential reason behind this interesting finding. We will present more results in Section~\ref{subsub:results:subjective:advantage-adaptive}.

\subsection{App Usage Behavior}\label{subsec:results_appusage_behavior}
\label{sub:results:usage}
In addition to intervention accuracy and receptivity, app usage behavior patterns were also important metrics to objectively measure the impact of interventions.
We analyzed the app usage logs to investigate the changes in participants' app usage frequency (Section~\ref{subsub:results:usage:frequency}) and duration (Section~\ref{subsub:results:usage:duration}) between the baseline week and the intervention phase.
Similar to Section~\ref{sub:results:intervention}, we also normalized our data against the baseline week data to reduce the bias introduced by individual differences.

\begin{figure}[b]
    \centering
    \begin{minipage}[b]{0.42\textwidth}
        \centering
        \includegraphics[width=\textwidth]{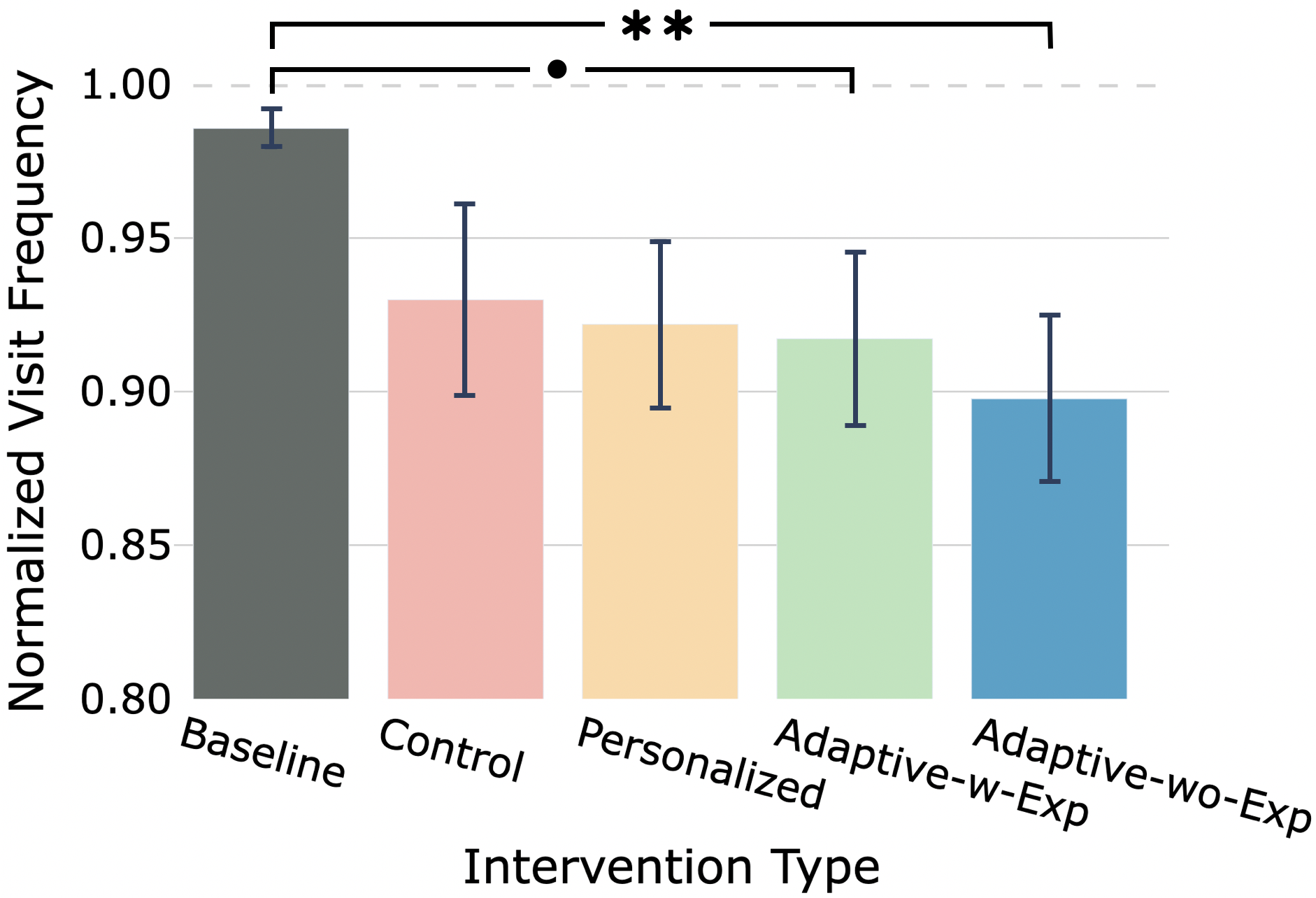}
    \end{minipage}
    \hspace{0.1cm}
    \begin{minipage}[b]{0.42\textwidth}
        \centering
        \includegraphics[width=\textwidth]{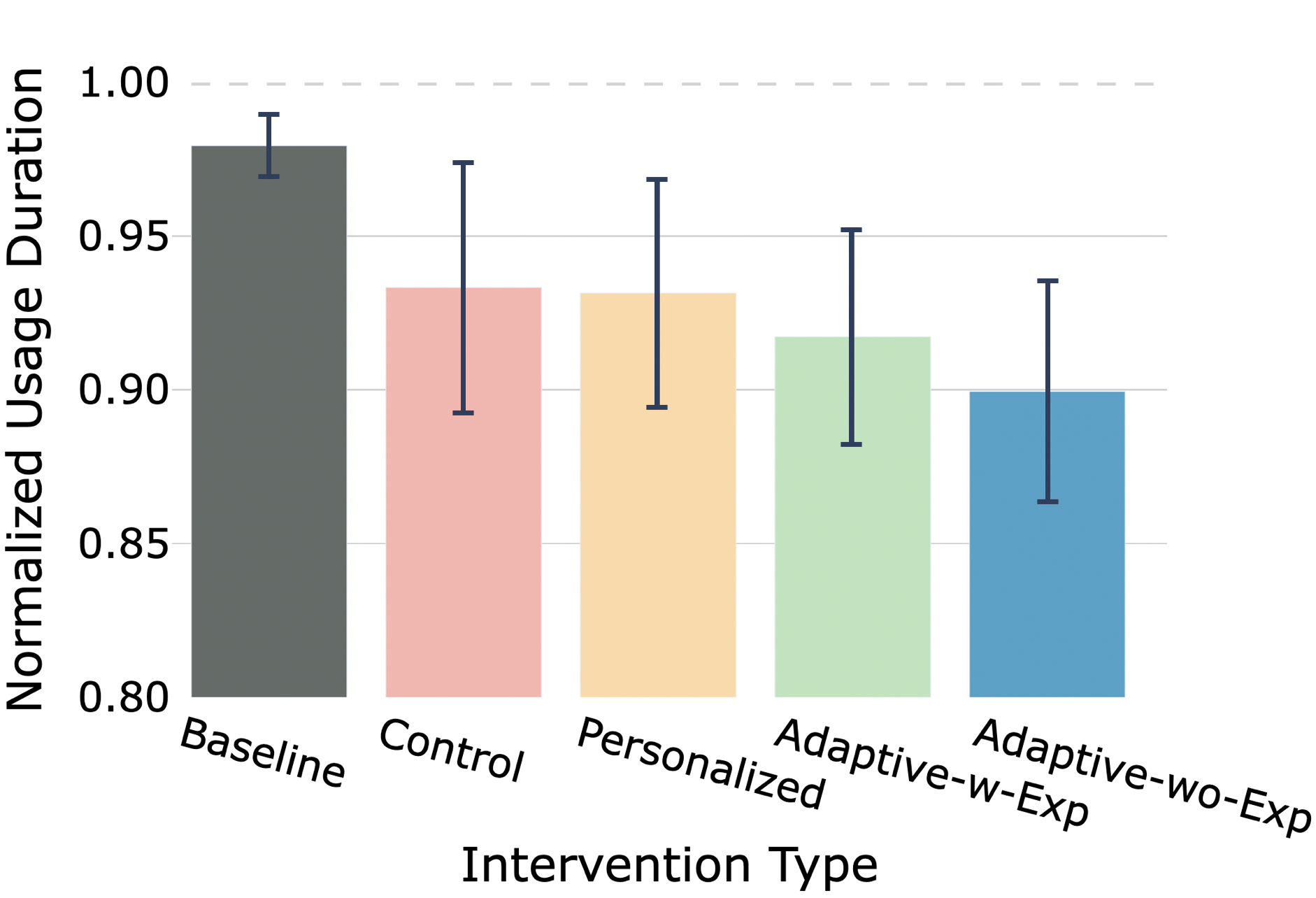}
    \end{minipage}
    \caption{App usage visit frequency (Top) and usage duration (Bottom). The two adaptive methods reduced the most app visit frequency and usage duration. Interestingly, in contrast to Figure~\ref{fig-intv-acc-and-rec-agg-3types}-\ref{fig:receptivity_overtime}, showing explanations did not augment the performance from the perspective of app usage behavior.}
    \label{fig:usage-visit-freq-spent-time}
\end{figure}

\subsubsection{Change of App Visit Frequency}
\label{subsub:results:usage:frequency}
\textbf{The two adaptive methods achieved a significant or marginally significant reduction in visit frequency compared to the base week. However, showing explanations was not as helpful.}
Figure~\ref{fig:usage-visit-freq-spent-time} (Left) compares the normalized visit frequency of the four intervention types.
The average daily visit frequency to monitored apps during the baseline collection period was 94.97 times~(SD=52.57).
Our results indicated that the visit frequency was reduced for all intervention types: \textit{Control} (93.0\%), \textit{Personalized}~(92.2\%), \textit{Adaptive-w-Exp}~(91.7\%), and \textit{Adaptive-wo-Exp}~(89.8\%).
We ran a GLMM on the visit frequency, with the intervention type as the main effect and participant ID as the random effect, which showed significance ($\chi^2(4)$=13.85, $p$<0.01).
Post-hoc results with Holm-Bonferroni correction showed that the visit frequency during the days of \textit{Adaptive-wo-Exp} intervention was significantly lower than the baseline week ($p$<0.01), and that the frequency of \textit{Adaptive-w-Exp} show marginal significance ($p$=0.07<0.1). This showed the advantage of the two adaptive methods over the \textit{Personalized} and \textit{Control} methods.
However, although the direct comparison between \textit{Adaptive-wo-Exp} and \textit{Adaptive-w-Exp} was not significant, we observed an interesting reversed effect of explanations: In Sections~\ref{sub:results:intervention} and \ref{sub:results:usage}, explanations could help to improve the intervention accuracy and receptivity; However, when looking into the app visit frequency, adaptive intervention without explanations had better performance.

\subsubsection{Change of App Usage Duration}
\label{subsub:results:usage:duration}
We also observed similar trends for app usage duration.
The average time spent on monitored apps during the baseline week was 214.00 minutes (SD=103.57).
The usage duration was reduced for all intervention types: \textit{Control} (93.3\%), \textit{Personalized}~(93.1\%), \textit{Adaptive-w-Exp}~(91.7\%), and \textit{Adaptive-wo-Exp}~(89.9\%). We still observe the similar advantage of \textit{Adaptive-wo-Exp} over \textit{Adaptive-w-Exp}, but the GLMM on usage duration did not show significance ($\chi^2(4)$=2.62, $p$=0.62).
With \textit{Adaptive-w-Exp}, although participants recognized and stopped more immediate overuse behavior, their overall usage patterns did not change much as \textit{Adaptive-wo-Exp}. We discuss these findings more in Section~\ref{sub:results:subjective} and Section~\ref{sub:discussion:mixed_effect}.

\subsection{Subjective Measure}
\label{sub:results:subjective}
In addition to the intervention accuracy, receptivity, and app usage behavior results, participants' survey responses and comments during exit interviews also provided interesting insights.

\begin{figure*}[b]
    \centering
    \begin{subfigure}[b]{152pt}
        \centering
        \includegraphics[width=152pt]{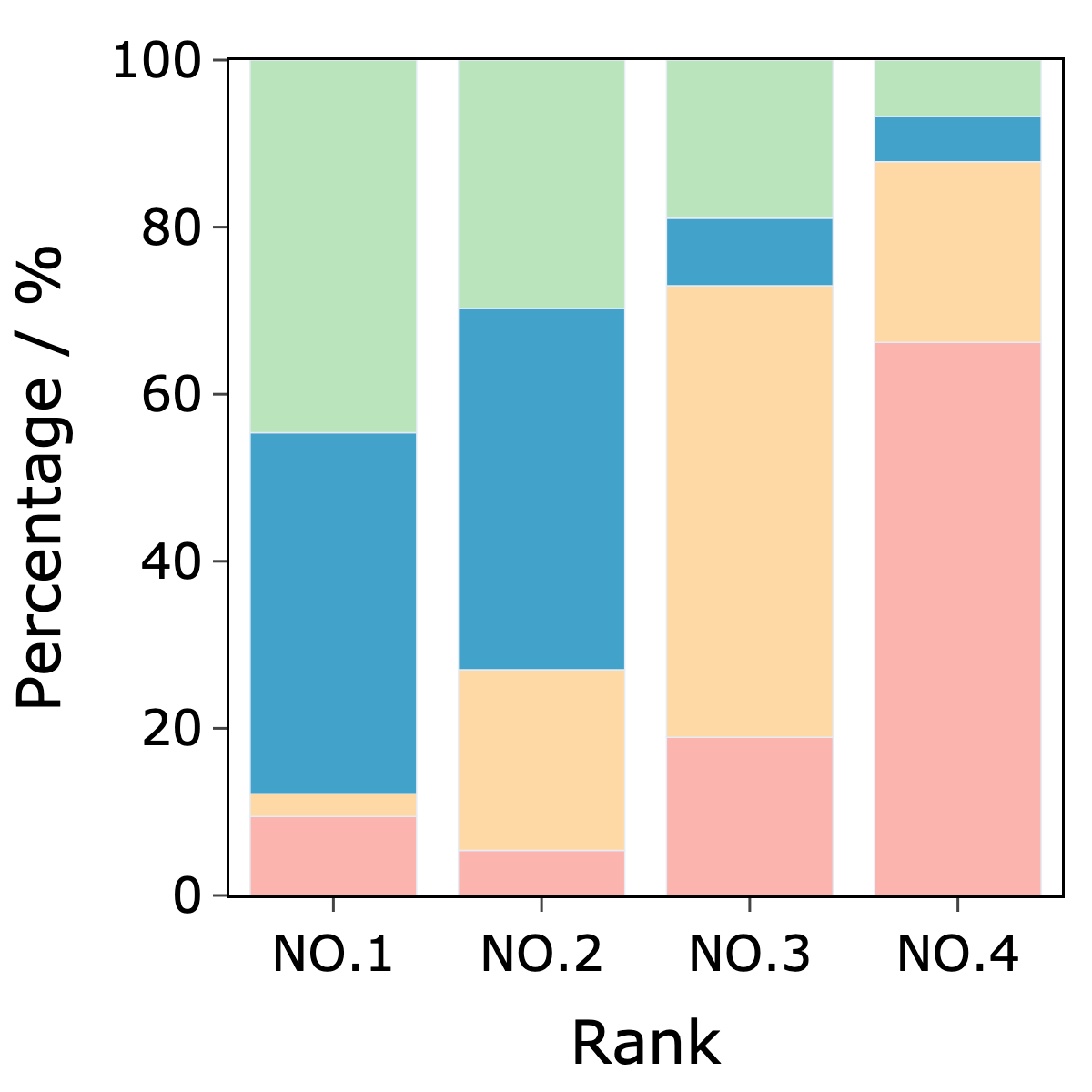}
    \end{subfigure}%
    \hfill
    \begin{subfigure}[b]{354.2pt}
        \centering
        \captionsetup{width=405pt}
        \includegraphics[width=354.2pt]{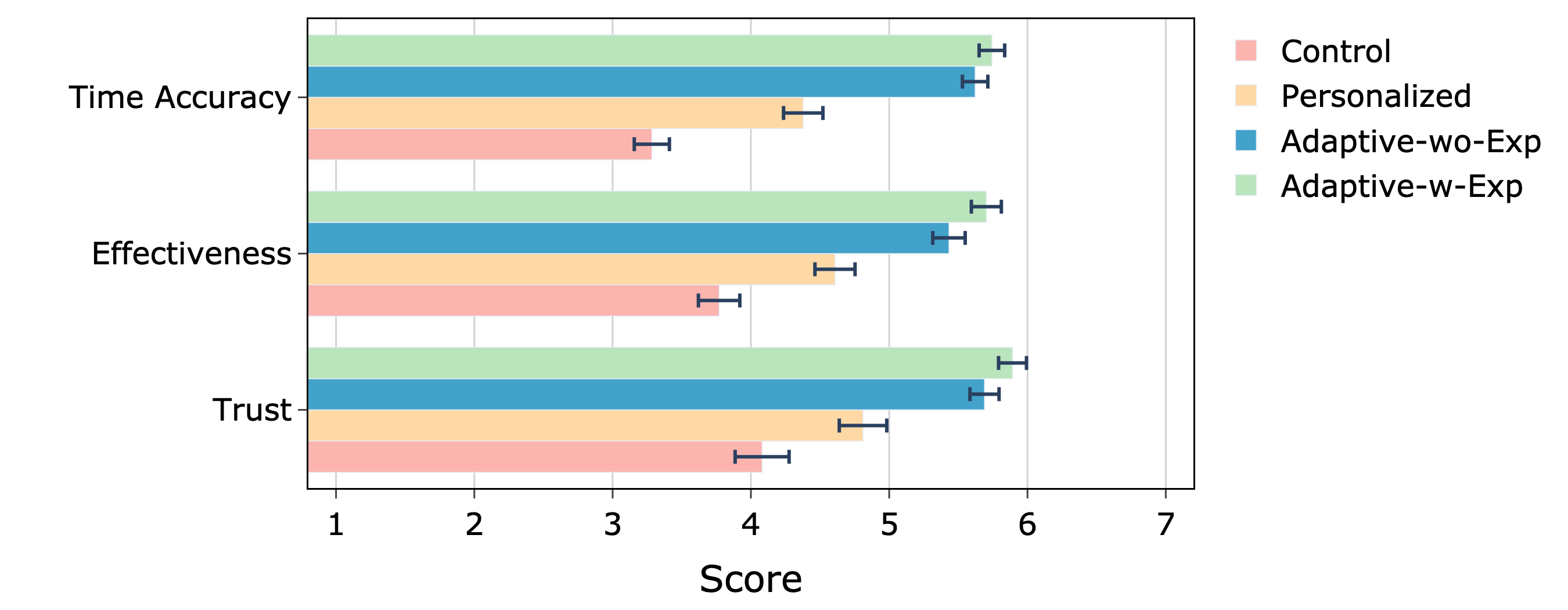}
    \end{subfigure}
    \caption{Survey Results Summary. (Left) User Preference Rankings among The Four Intervention Methods. (Right) User Ratings on Intervention Time Accuracy, Perceived Effectiveness, and Level of Trust to Different Methods. The two adaptive methods received the highest user subjective preference and ratings.}
    \label{fig:survey}
\end{figure*}

\subsubsection{Clear Advantage of Adaptive Intervention Methods}
\label{subsub:results:subjective:advantage-adaptive}

Overall, participants had a clear preference for \textit{Adaptive-w-Exp} (\ie \projectname{}) and \textit{Adaptive-wo-Exp}, followed by \textit{Personalized}, and then \textit{Control}. The left of Figure~\ref{fig:survey} presents participants' ranking results among the four intervention techniques. \textit{Adaptive-w-Exp} received the most NO.1 ranking (45\% of participants), and \textit{Adaptive-wo-Exp} came as the second (43\%).
This observation was confirmed by a non-parametric Friedman test on ranking numbers that showed strong significance ($\chi^2(3)$=88.01, $p$<0.001). Our post-hoc pairwise comparison (Wilcoxon signed-rank test with Holm-Bonferroni correction) indicated significance among all pairs ($p$s<0.001) except \textit{Adaptive-w-Exp} \vs \textit{Adaptive-wo-Exp} ($p$=0.45).

Meanwhile, participants' ratings on the time accuracy, intervention effectiveness, and level of trust were consistent with the ranking results, as shown in the right of Figure~\ref{fig:survey}. We ran three individual Friedman tests on the three metrics. All of them indicated significance ($p$s<0.001). The post-hoc analysis showed that almost all pair comparisons were significant (for \textit{Adaptive-w-Exp} \vs \textit{Adaptive-wo-Exp}: $p_{\textit{effectiveness}}$<0.01, $p_{\textit{trust}}$<0.05, all others $p$s<0.001). The only exception was \textit{Adaptive-w-Exp} \vs \textit{Adaptive-wo-Exp} on time accuracy ($p$=0.15).

Our interview data also triangulated these quantitative findings. Many participants felt the difference when comparing \textit{Personalized} and \textit{Control}. \pquote{14}{The random version [Control] didn't make sense, and the timing was strange some days. I think the personalized ML version [Personalized] was consistent with my annotations a few weeks ago.} A similar distinction was also observed when comparing the two adaptive versions and \textit{Personalized}. \pquote{34}{I can feel that the adaptive version [Adaptive-wo-Exp] has been learning about my behavior. At the later stage of the study, some days more interventions would pop up if I overused more.} \pquote{55}{The version with explanations [Adaptive-w-Exp] is clearly adaptive. The intervention timing became more comfortable after I used it for a while.}
It is noteworthy that the interface of \textit{Adaptive-wo-Exp}, \textit{Personalized}, and \textit{Control} were the same, and participants only learned the exact dates for intervention methods after the study finished. So, their feeling of differences was mainly based on their experience of the intervention timing.

These findings are in line with the results in the previous section about the advantage of \textit{Personalized} over \textit{Control}, and more importantly, the advantage of \textit{Adaptive-w-Exp} and \textit{Adaptive-wo-Exp} over other two methods.

Moreover, we also noticed that there was a small proportion of users ranking \textit{Control} as the top 1 type (Figure~\ref{fig:survey} left). Participants commented that this technique was ``\textit{surprising}/\textit{unexpected}''. This was in contrast to the \textit{Personalized} method. \pquote{15}{Later in the study, I could somehow expect when it [Personalized] would show up. But that method [Control] is hard to predict. So sometimes it is refreshing.}
This is also supported by previous work~\cite{kovacs2018rotating}, which could explain the increasing trend of \textit{Control}'s receptivity and the decreasing trend of \textit{Personalized} over time in Section~\ref{subsub:results:intervention:receptivity-time} and Figure~\ref{fig:receptivity_overtime}.

\subsubsection{Trade-off between with \vs without Explanation}
\label{subsub:results:subjective:with-without-exp}
We also had interesting observations that could explain the difference between intervention receptivity (where \textit{Adaptive-w-Exp} had the best \nobreak performance, as shown in Figure~\ref{fig:intv-acc-and-rec-agg-4types} and \ref{fig:receptivity_overtime}) and app usage behavior (where \textit{Adaptive-wo-Exp} was the best, Figure~\ref{fig:usage-visit-freq-spent-time}).

Our survey results suggested that \textit{Adaptive-w-Exp} and \textit{Adaptive-wo-Exp} had similar performance. We also found diversity in preference ranking of \textit{Adaptive-w-Exp}: Although \textit{Adaptive-w-Exp} received 45\% of the NO.1 voting (compared to a similar 43\% for \textit{Adaptive-wo-Exp}), it also received 19\% of the NO.3 voting (compared to a much lower 8\% for \textit{Adaptive-wo-Exp}). While most participants liked \textit{Adaptive-w-Exp}, a certain proportion of participants found it less preferable.

We dug deep into this difference during our exit interviews.
On the one hand, participants who preferred \textit{Adaptive-w-Exp} found explanations could trigger more self-awareness: \pquote{26}{Seeing the explanations could help me to better self-reflect, which often made me stop using my phone.} \pquote{29}{Those explanations pushed me to think more about the reason behind my phone usage.} \pquote{24}{Explanations helped me to trust the system better.} These results indicated that showing explanations could better trigger System 2 (the reasoning and analytical system) with reasoning and self-analysis and improve users' trust in the intervention.
These could explain the significantly better effectiveness and level of trust in \textit{Adaptive-w-Exp} (yet the effective sizes were limited $r_{\textit{effectiveness}}$=0.20, $r_{trust}$=0.16). 
On the other hand, participants who did not like \textit{Adaptive-w-Exp} found explanations overly broad and sometimes confusing. \pquote{34}{Sometimes, the explanations felt accurate. But they were very broad so I am not sure.} Some participants found explanations unnecessary. \pquote{59}{I was aware of my phone overuse, so I didn't need explanations.}
These results suggest a more detailed and fine-grained explanation could be helpful for smartphone overuse intervention.

These diverse user reactions toward intervention explanations could explain the mixed results when comparing \textit{Adaptive-w-Exp} and \textit{Adaptive-wo-Exp}.
We will have more discussion about this in Section~\ref{sec:discussion}.

\subsection{Summary of Results}
\label{sub:results:summary}
Our 8-week field experiment showed that AI-powered JITAI interventions effectively reduce smartphone overuse. Our two \textit{Adaptive} models provided significantly more accurate interventions compared to \textit{Control} (55.5\%) and \textit{Personalized}~(32.8\%) groups. This trend was consistent for intervention receptivity: participants were significantly more receptive to the two \textit{Adaptive} models compared to \textit{Personalized} (8.0\%) and \textit{Control} (29.0\%) models. Furthermore, the intervention accuracy and receptivity were further enhanced with explanations.
\textit{Adaptive-w-Exp}, \ie our complete system \projectname{}, could significantly better help users to recognize their overuse (high accuracy) than \textit{Adaptive-wo-Exp} (53.8\%), \textit{Personalized} (66.9\%), and \textit{Control} (97.5\%)
methods.
Similarly, explanations helped users to be more receptive to interventions and quit using apps. \textit{Adaptive-w-Exp} was more receptive than \textit{Adaptive-wo-Exp} (11.4\%), \textit{Personalized} (18.9\%), and \textit{Control} (39.6\%).
We also discovered that the receptivity of adaptive models improved throughout the intervention period, showing the potential of benefiting from long-term deployment with adaptive ML models.

Regarding the actual smartphone usage behavior, all intervention types helped users reduce usage compared to the baseline week.
We observe a significant reduction in app visit frequency for \textit{Adaptive-wo-Exp} (8.9\%) and a marginally significant reduction for \textit{Adaptive-w-Exp} (7.0\%).
Analysis of subjective responses also aligned with our quantitative findings. Most participants ranked \textit{Adaptive-w-Exp} and \textit{Adaptive-wo-Exp} as their preferred options. Moreover, time accuracy, effectiveness, and trust were consistent with the results by showing the superiority of the two adaptive models. Interestingly, we observed an unexpected mixed effect of explanations.
The intervention accuracy and receptivity results indicated the advantage of explanations, while the app usage behavior suggested the opposite. Our qualitative results revealed that some users appreciated explanations for higher transparency and trustworthiness. On the other hand, other participants found explanations sometimes redundant or overly broad. We discuss this interesting observation in the next section.

\section{Discussion}
\label{sec:discussion}
We designed and developed a novel ML-based explainable JITAI system \projectname{} for smartphone overuse intervention. To systematically evaluate the effectiveness of making the system adaptive and explainable, we conducted a micro-randomized study to deploy and measure four different intervention types. Each type added one more component on top of the previous version: ML-based intelligence (\textit{Personalized} \vs \textit{Control}), adaptivity (\textit{Adaptive-wo-Exp} \vs \textit{Personalized}), and explainability  (\textit{Adaptive-w-Exp} \vs \textit{Adaptive-wo-Exp}).
Our results demonstrate each component can improve the performance of the intervention system to some extent, with an interesting observation of the mixed effect of explanations.
Combining these components, \projectname{} provides a trustworthy and effective intervention with accurate timing while adapting to individuals' behaviors.
In this section, we discuss the potential reasons behind the explanations' effect (Sec.~\ref{sub:discussion:mixed_effect}), the design considerations and takeaways from our field experiment (Sec.~\ref{sub:discussion:design_consideration}), the ethical concerns accompanying AI-based JITAI systems (Sec.~\ref{sub:discussion:ethical_concerns}), as well as the limitations in our work (Sec.~\ref{sub:discussion:limitation}).

\subsection{The Mixed Effect of Explanations}
\label{sub:discussion:mixed_effect}
In our field experiment, the advantages of \textit{Personalized} over \textit{Control} and \textit{Adaptive-wo-Exp} over \textit{Personalized} are stable across different metrics. However, the comparison between \textit{Adaptive-wo-Exp}  and \textit{Adaptive-w-Exp} shows diverse results.
In Figures~\ref{fig:intv-acc-and-rec-agg-4types} and \ref{fig:receptivity_overtime}, the advantage of \textit{Adaptive-w-Exp} is clear, while in Figure~\ref{fig:usage-visit-freq-spent-time} we observe the advantage of \textit{Adaptive-w-Exp} instead.
These results indicate that during the experiment, participants tended to agree with the intervention timing (higher accuracy) and leave the target apps (higher receptivity) when interventions provided explanations. However, such ``successful intervention'' did not have a lasting effect. Participants still visited and spent more time in target apps.
Although our explanations successfully triggered their System 2 and led to usage pauses, some participants did not effectively internalize the explanation content and were ``pushed'' by explainable interventions without deep reflection.
Our interview results also support this. Although most participants found explanations helpful for self-reflection, some found explanations confusing and overly broad.

This illustrates the need for more advanced explanation generation techniques in future deployment. 
Now that we have built adaptive ML models, future explanations should be dynamic, personalized, and adaptive to users. Our interview results reveal that individuals have different preferences in the level of detail. Thus, our system must adapt explanations to fulfill users' specific needs~\cite{thomson2020knowledge}.
Meanwhile, recent research suggests a few promising directions, such as explanation selection (to ensure preference alignment)~\cite{lai2023selective} and verifiability (to verify the correctness of AI outputs)~\cite{fok2023search}.
Future work can be explored along with these directions to enhance the effectiveness of explanations further.

\subsection{Design Consideration of Intelligent JITAI with Human-in-the-Loop}
\label{sub:discussion:design_consideration}

We made a range of design decisions in our deployment. We reflect on important considerations and share the lessons from our study.

\textbf{Alternative User-in-the-Loop Labels.}
In our study, we designed a simple single-click feedback button to collect user feedback and establish the human-AI loop (see Figure~\ref{fig:intervention-interface}). We then used such feedback as new labels to tune our ML model. This design has pros and cons. On the one hand, it retrieves users' real-time reactions explicitly so that the model adapts toward users' subjective experience and preference, providing transparency and user agency to some extent. On the other hand, it requires extra effort from users and can miss data when users do not provide feedback.
Moreover, this approach does not consider potential bias (compulsively engaged users providing incorrect labels). 
Another alternative is to leverage users' reactions towards the intervention as implicit feedback labels (\eg leaving the app could be marked as being receptive to the intervention). This method is also adopted by some previous work in AI-based JITAI systems~\cite{Rabbi2015,liao_personalized_2020}. It reduces user effort and adapts the model based on their actual behaviors. However, such implicit labels can be affected by noisy behavior, \ie users could leave/stay in the app for external reasons other than the intervention. Additionally, a hybrid method utilizing both explicit (user-provided) and implicit (behavioral reactions) labels could be implemented. This hybrid model could be refined by measuring the consistency between implicit and explicit labels to assign varying weights to samples, facilitating the enhancement or updating of the model. Moreover, involving health experts in the human-AI loop could provide a valuable solution. Collaborating with experts allows for a more nuanced and balanced definition of overuse, incorporating both user perspectives and health-related guidelines. This collaboration ensures a more objective and informed approach toward setting criteria that align with both user behavior and health considerations, thereby enhancing the accuracy and reliability of the model's updates and interventions.
Researchers, designers, and developers must carefully inspect the specific use cases and choose between explicit and implicit feedback or a combination of both.

\textbf{Real-time \vs Reflective Feedback.}
Other than collecting user feedback \textit{in situ} (\ie when using target apps), we also explored another method to ask users to label their data at the end of the day.
This post-hoc labeling offers users more time to reflect on their behavior. However, recalling earlier smartphone usage cases can be challenging, especially for quick usage such as habitual phone checks.
We introduced this method in our experiment for participants who wanted to make up for missing labels. However, our results indicated that they barely used this method (around 3\%), thus we did not include them in our analysis. This was mainly because our label collection and feedback design was simple enough. A reflective feedback mechanism could be a promising solution for other behavior intervention studies involving a more complex label collection process.

\textbf{Prediction Model Update Methods and Frequency.}
We did not explore more advanced models in \projectname{}, such as deep learning or recent large language models~(LLMs)~\cite{mahmood2023llm, yang2023talk2care}, as the model itself is not the main focus of our paper.
\projectname{} employs re-training with recency-based weight assignment for model updates. Although this method is robust, other advanced methods, such as reinforcement learning, can be explored in future work.
Moreover, the update frequency of the prediction model is crucial for the system's adaptability.
We updated the model daily to balance our current design's performance and computational costs. But there can be other options. A high frequency of updates (\eg hourly or even after each interaction) can allow the system to rapidly adapt to users' changing behavior and provide more timely and relevant interventions. However, this comes with higher computational costs and the risk of overfitting to temporary changes in user behavior.
Conversely, a lower frequency of updates (\eg weekly or monthly) reduces the computational load and the risk of overfitting. Still, it may result in the system being slower to adapt to meaningful changes in user behavior.
There is a trade-off between adaptability and stability that must be carefully considered.
In addition, the trade-off is also impacted by specific applications. Interventions for mental health may require a different frequency than the ones for smartphone overuse.
Future work could explore adaptive update frequencies, where the model update frequency is dynamically adjusted based on the stability of user behavior and the model's performance.

\textbf{Handling ``Cold-Start'' in JITAI-based Interventions.}
In our study, we devoted the first two weeks to data collection before deploying the intervention. This could be hard to achieve in real-world scenarios. To address this ``cold-start'' challenge, one promising future approach involves unsupervised learning~\cite{ghahramani2003unsupervised, bengio2012deep} where users will not be required to provide labels. Instead, the model will grasp smartphone usage patterns by leveraging (\eg by clustering) users' historical data.
Another potential strategy involves few-shot domain adaptation~\cite{wang2020generalizing, gong2023dapper, snell2017prototypical,xu_enabling_2022, gong_metasense_2019}, where we can pre-train a model with a dataset (such as from this study) and then fine-tune the model with a small amount of additional data from new users. Additionally, test-time adaptation~\cite{gong2022note, gong2023sotta}, an advanced domain adaptation technique, could directly utilize test-time data to adapt a global model to a new user without requiring any collected training data.

\textbf{Dynamic Features for Longitudinal Model Deployment.}
In our study, we conducted feature selection using the first two weeks of the data and kept the feature set static throughout the experiment.
However, for long-term deployment, the importance of different features may change over time. Therefore, dynamic feature selection can be applied. One potential method is selecting each model update's most relevant feature set. This may help the model to have a better performance over time. However, similar to the frequent model updates discussed above, dynamic feature selection will introduce additional complexity and computational requirements. It may also result in a less stable model, which can be challenging for explanation generation and user trust building.
The same trade-off between adaptability and stability is also needed for dynamic feature selection.

\textbf{Explanation Level of Details.}
As mentioned in the previous section, our current design of high-level intervention explanations could be too general and confusing. Sec.~\ref{sub:discussion:mixed_effect} discusses the potential of personalized and adaptive explanation generation.
However, overly detailed explanations may inadvertently reveal sensitive information about user behavior, which can raise privacy concerns~\cite{kim_prediction_2022,barredo_arrieta_explainable_2020}.
It is an open research question on providing appropriate detail for model explanation.
For behavior change targets that are more objective (\eg smartphone overuse), providing more detailed explanations can be a good idea.
While for more abstract targets (\eg stress management), high-level and abstract explanations may be more appropriate~\cite{kim_prediction_2022}.

\subsection{Ethical Concerns and Risk of AI-based Intervention System}
\label{sub:discussion:ethical_concerns}
Despite the promising performance of our explainable JITAI, we also highlight the important ethical concerns of such an intelligent intervention system. These concerns must be addressed before any real-world, large-scale deployment.
First, there is the risk of wrongly predicting smartphone overuse. Our best performance achieved an F1 score of 67\%. It may occasionally make incorrect predictions and lead to poorly timed interventions, which are annoying or even harmful to users. For example, a false positive that incorrectly identifies a user as overusing their smartphone when they are using it for an important task may cause unnecessary stress and disrupt their workflow.
Similarly, due to the limitations of the explanation method and model performance, the explanation content may not accurately reflect the actual reasons. These wrong explanations can lead to confusion and mistrust of the system and may result in users ignoring or rejecting the interventions.
Therefore, it is essential to carefully evaluate the model prediction performance and explanation quality. Other than exploring more advanced ML algorithms,  LLMs~\cite{bubeck_sparks_2023,li2022explanations,xu2023leveraging,wu2024mindshift} may offer a new method to generate appropriate and convincing explanation content.

Besides, when an intervention is personalized, users, especially younger ones, could be biased towards being more receptive to adopting it~\cite{guo2016privacy}. Therefore, a misalignment between subjective measures (users' feedback) and objective measures (their actual behavior) in the study could exist. Our discussion about alternative user-in-the-loop labels in Section~\ref{sub:discussion:design_consideration} could be a potential solution. This factor should be carefully considered when deploying interventions with explicit personalized components.

Meanwhile, privacy is another critical concern. Our current system adopts a  centralized learning method that merges all users' data for model training. Future work can explore edge computing methods such as federated learning~\cite{li_federated_2020} to address privacy concerns.

\subsection{Limitations}
\label{sub:discussion:limitation}
There are a few limitations in our work.
First, we mainly focused on young adults. Our study population had a limited age range; thus, the findings of our results may not be generalizable to other population groups. Meanwhile, there is a lack of exploration on the fairness evaluation of our methods.
Second, our micro-randomized study design took the daily level as the randomization unit. This enabled us to conduct a within-subject design within a feasible period. However, we could not investigate the lasting effect of different intervention methods as they were mixed. Meanwhile, our observational study may neglect unknown confounding variables beyond this paper's scope. 
For instance, while we employed a time-based train-test split to train the base model, we acknowledge that mitigating the `observational effect' (the impact of monitoring and labeling) might pose a challenge. Besides, although we revealed the exact dates of different interventions during exit interviews and surveys, participants' responses might still be inaccurate or biased by their memory. Third, we updated intelligent adaptive models at midnight, which might not align with college students' sleeping schedules.
Last, our work only considered smartphone overuse as a general intervention target. The specific types of overuse, such as excessive use of social media or video gaming, were not investigated in detail in this study. Similarly, as mentioned earlier, we didn't experiment with more intervention methods other than digit typing, as the specific intervention method is not the focus of our work.
\section{Conclusion}
\label{sec:conclusion}
In this paper, we propose a novel AI-powered explainable JITAI system, \projectname{}, for smartphone overuse intervention.
Our system captures user context and behavior, leverages AI to infer smartphone overuse scenarios, introduces interventions when overuse is detected, provides explanations, and updates the intervention model iteratively based on human-in-the-loop feedback to form a human-AI loop.
In order to measure the effectiveness of \projectname{}, we conducted an 8-week field experiment (N=71) and compared four intervention types.

Our results not only showed the advantage of the ML component (the static ML-powered version over the basic control), but more importantly, underscored the advantages of adaptive intervention types compared to the static version, with significantly better intervention accuracy (32.8\%) and receptivity (8.0\%).
Furthermore, including explanations in our system significantly amplified its accuracy (53.8\%) and receptivity (11.4\%).
In addition, users exhibited reduced visit frequency to apps they considered unproductive when engaged with adaptive models (7.0-8.9\%).
Findings from our qualitative analysis echoed the quantitative results, with users expressing a clear preference for adaptive interventions. We also observed an interesting mixed effect of explanations, which could shed light on future research direction.
We further highlighted the important ethical concerns of AI-based intervention systems for real-world deployment.
We envision our work can be applied beyond the field of smartphone overuse and inspire future practitioners to explore more advanced intervention techniques with a human-AI loop.

\begin{acks}
We express our gratitude to all the participants who contributed to our longitudinal user study. We extend our appreciation to Jennifer Mankoff for her insights and discussions that significantly enriched this project, and to Zihan Yan for her assistance in the pilot study. This paper is based upon work supported by the VW Foundation, Quanta Computing, the Natural Science Foundation of China (NSFC) under Grant Number 62132010, Young Elite Scientists Sponsorship Program by CAST under Grant Number 2021QNRC001, Tsinghua University Initiative Scientific Research Program and Institute of Information \& communications Technology Planning \& Evaluation (IITP) under Grant Number 2022-0-00495.
\end{acks}

\bibliographystyle{ACM-Reference-Format}
\bibliography{main}

\appendix
\onecolumn
\appendix

\section{Explanation Examples} 
\label{sec:appendix:explanation}

\aptLtoX[graphic=no,type=html]{
\begin{table}[h]
\vspace{0.3cm}
\caption{Examples of Feature Explanations at Different Explanation Levels.}
\label{tab:feature-explanation}
\resizebox{1\linewidth}{!}{
\begin{tabular}{c|c|cc}
\hline \hline
\multicolumn{1}{c|}{\multirow{2}{*}{\begin{tabular}[c]{@{}c@{}}\textbf{Model Feature} \end{tabular}}} & \multicolumn{1}{c|}{\multirow{2}{*}{\begin{tabular}[c]{@{}c@{}}\textbf{Readable Name} \end{tabular}}}  & \multicolumn{2}{c}{\textbf{Explanation}}                                         \\ \cline{3-4} 
& & \multicolumn{1}{c|}{\textbf{High-level}} & \textbf{Low-level} \\ \hline

\textit{numViewScrolledCurrentAppCategory}                                                                                           & \multicolumn{1}{c|}{Number of Scrolls in Current App Category}       & \multicolumn{1}{c|}{Phone \& App Use}       & Number of Interactions         \\ 

\textit{sumDurationDischarge}                                                                                      & \multicolumn{1}{c|}{Battery Discharge Duration}      & \multicolumn{1}{c|}{Phone \& App Use}       & Battery Usage          \\\hline

\textit{durationMobile}                                                                                           & \multicolumn{1}{c|}{Duration of Being Mobile}       & \multicolumn{1}{c|}{Activity}       & Duration of Being Mobile         \\ 

\textit{avgLux}                                                                                      & \multicolumn{1}{c|}{Average Lux in Light Conditions}      & \multicolumn{1}{c|}{Activity}       & Light Conditions        \\\hline

\textit{countScansMostFrequentDevice}                                                                                           & \multicolumn{1}{c|}{Number of Frequently Scanned Devices}       & \multicolumn{1}{c|}{Social}       & Number of Nearby Devices        \\ 

\textit{timeFirstSent}                                                                                      & \multicolumn{1}{c|}{Time of First Sent Message}      & \multicolumn{1}{c|}{Social}       & Time of Sent Message   \\\hline

\textit{timeAtTopOneLocation}                                                                                           & \multicolumn{1}{c|}{Time Spent at Top One Location}       & \multicolumn{1}{c|}{Location}       & Time at Frequent Locations        \\ 

\textit{minLengthStayAtClusters}                                                                                      & \multicolumn{1}{c|}{Minimum Stay at Frequent Locations}      & \multicolumn{1}{c|}{Location}       & Time at Frequent Locations        \\\hline

\textit{isNight}                                                                                           & \multicolumn{1}{c|}{Whether it is the Night Time}       & \multicolumn{1}{c|}{Time}       & the Night Time         \\

\hline \hline
\end{tabular}
}
\end{table}
}{
\begin{table}[h]
\vspace{0.3cm}
\caption{Examples of Feature Explanations at Different Explanation Levels.}
\label{tab:feature-explanation}
\resizebox{1\linewidth}{!}{
\begin{tabular}{c|c|cc}
\hline \hline
\multicolumn{1}{c|}{\multirow{2}{*}{\begin{tabular}[c]{@{}c@{}}\textbf{Model Feature} \end{tabular}}} & \multicolumn{1}{c|}{\multirow{2}{*}{\begin{tabular}[c]{@{}c@{}}\textbf{Readable Name} \end{tabular}}}  & \multicolumn{2}{c}{\textbf{Explanation}}                                         \\ \cline{3-4} 
& & \multicolumn{1}{c|}{\textbf{High-level}} & \textbf{Low-level} \\ \hline

\textit{numViewScrolledCurrentAppCategory}                                                                                           & \multicolumn{1}{c|}{Number of Scrolls in Current App Category}       & \multicolumn{1}{c|}{Phone \& App Use}       & Number of Interactions         \\ 

\textit{sumDurationDischarge}                                                                                      & \multicolumn{1}{c|}{Battery Discharge Duration}      & \multicolumn{1}{c|}{Phone \& App Use}       & Battery Usage          \\\cdashline{1-4}

\textit{durationMobile}                                                                                           & \multicolumn{1}{c|}{Duration of Being Mobile}       & \multicolumn{1}{c|}{Activity}       & Duration of Being Mobile         \\ 

\textit{avgLux}                                                                                      & \multicolumn{1}{c|}{Average Lux in Light Conditions}      & \multicolumn{1}{c|}{Activity}       & Light Conditions        \\\cdashline{1-4}

\textit{countScansMostFrequentDevice}                                                                                           & \multicolumn{1}{c|}{Number of Frequently Scanned Devices}       & \multicolumn{1}{c|}{Social}       & Number of Nearby Devices        \\ 

\textit{timeFirstSent}                                                                                      & \multicolumn{1}{c|}{Time of First Sent Message}      & \multicolumn{1}{c|}{Social}       & Time of Sent Message   \\\cdashline{1-4}

\textit{timeAtTopOneLocation}                                                                                           & \multicolumn{1}{c|}{Time Spent at Top One Location}       & \multicolumn{1}{c|}{Location}       & Time at Frequent Locations        \\ 

\textit{minLengthStayAtClusters}                                                                                      & \multicolumn{1}{c|}{Minimum Stay at Frequent Locations}      & \multicolumn{1}{c|}{Location}       & Time at Frequent Locations        \\\cdashline{1-4}

\textit{isNight}                                                                                           & \multicolumn{1}{c|}{Whether it is the Night Time}       & \multicolumn{1}{c|}{Time}       & the Night Time         \\

\hline \hline
\end{tabular}
}
\end{table}
}

\end{document}